\documentstyle[epsfig,graphics]{mn}

\begin{document}

\newcommand{\Zsolar}{\mbox{\,$\rm Z_{\odot}$}}
\newcommand{\Msolar}{\mbox{\,$\rm M_{\odot}$}}
\newcommand{\Lsolar}{\mbox{\,$\rm L_{\odot}$}}
\newcommand{\etal}{{et al.}\ }
\newcommand{\ang}{\mbox{$\rm \AA$}}
\newcommand{\xs}{$\chi^{2}$}
\newcommand{\xmm}{{\it XMM-Newton}}
\newcommand{\chandra}{{\it Chandra}}
\newcommand{\ls}{{\tiny \( \stackrel{<}{\sim}\)}}
\newcommand{\gs}{{\tiny \( \stackrel{>}{\sim}\)}}
\newcommand{\asec}{$^{\prime\prime}$}
\newcommand{\amin}{$^{\prime}$}
\newcommand{\lx}{L$_{X}$}

\title[{\it XMM-Newton} observations of the late-stage merger-remnant galaxies NGC 3921 and NGC 7252]{{\it XMM-Newton} observations of the merger-remnant galaxies NGC 3921 and NGC 7252}
\author[L.A. Nolan \etal.]
{L.A. Nolan$^{1}$, T.J. Ponman$^{1}$, A.M. Read$^{2}$ $\&$ Fran\c cois Schweizer$^{3}$
\\
$^{1}$School of Physics and Astronomy, University of Birmingham, Birmingham B15 2TT\\
$^{2}$Department of Physics and Astronomy, University of Leicester, University Road, Leicester LE1 7RH\\
$^{3}$Carnegie Observatories, 813 Santa Barbara Street, Pasadena, CA 91101}

\date{Submitted for publication in MNRAS}

\maketitle
  
\begin{abstract}

Using the high sensitivity of \xmm, we have studied the X-ray
emission of the two proto-typical late-stage merger remnants, NGC 3921
and NGC 7252.  In the case of NGC 7252, this is complemented by
archival \chandra\ data.  We investigate the nature of the discrete
X-ray point source populations and the hot diffuse gas components in these
two galaxies, and compare them in the light of their different merger
ages and histories.

We detect 3 candidate ultra-luminous X-ray point sources in NGC 3921
and at least 6 in NGC 7252, for which we have high spatial resolution
\chandra\ data. These have luminosities ranging from
$\sim$1.4x10$^{39}-10^{40}$ erg s$^{-1}$ (for H$_{0}=$ 75 km s$^{-1}$
Mpc$^{-1}$). We expect these ULXs to be high mass X-ray
binaries, associated with the recent star formation in these two
galaxies.

Extended hot gas is observed in both galaxies. We have sufficient
counts in the \xmm\ data to fit two-component hot plasma models to
their X-ray spectra and estimate the X-ray luminosities of the hot
diffuse gas components to be 2.75x10$^{40}$ erg s$^{-1}$ and
2.09x10$^{40}$ erg s$^{-1}$ in NGC 3921 and NGC 7252,
respectively. These luminosities are low compared with the
luminosities observed in typical mature elliptical galaxies (\lx\
$\sim$10$^{41-42}$ erg s$^{-1}$), into which these merger remnants are
expected to evolve. We do not see evidence that the X-ray halos of
these galaxies are currently being regenerated to the masses and
luminosities seen in typical elliptical galaxies. The mass of atomic
gas available to fall back into the main bodies of these galaxies and
shock-heat to X-ray temperatures is insufficient for this to be the
sole halo regeneration mechanism. We conclude that halo regeneration
is most likely a long-term ($>$10 Gyr) process, occurring
predominantly via mass loss from evolving stars, in a sub-sonic
outflow stage commencing $\sim$ 2 Gyr after the merging event.

\end{abstract}

\begin{keywords}
galaxies: individual: NGC 3921 - galaxies: individual: NGC 7252 - galaxies: evolution - X-rays: galaxies - X-rays: ISM - X-rays:binaries
\end{keywords}

\section{Introduction}

The hypothesis that elliptical galaxies may be formed from the merging
of two disk galaxies was first proposed by Toomre \& Toomre (1972). It
is now known to be an essential part of galaxy formation, but there are
few merging systems close enough to be studied in depth.

\begin{figure*}

\raggedright{\epsfig{file=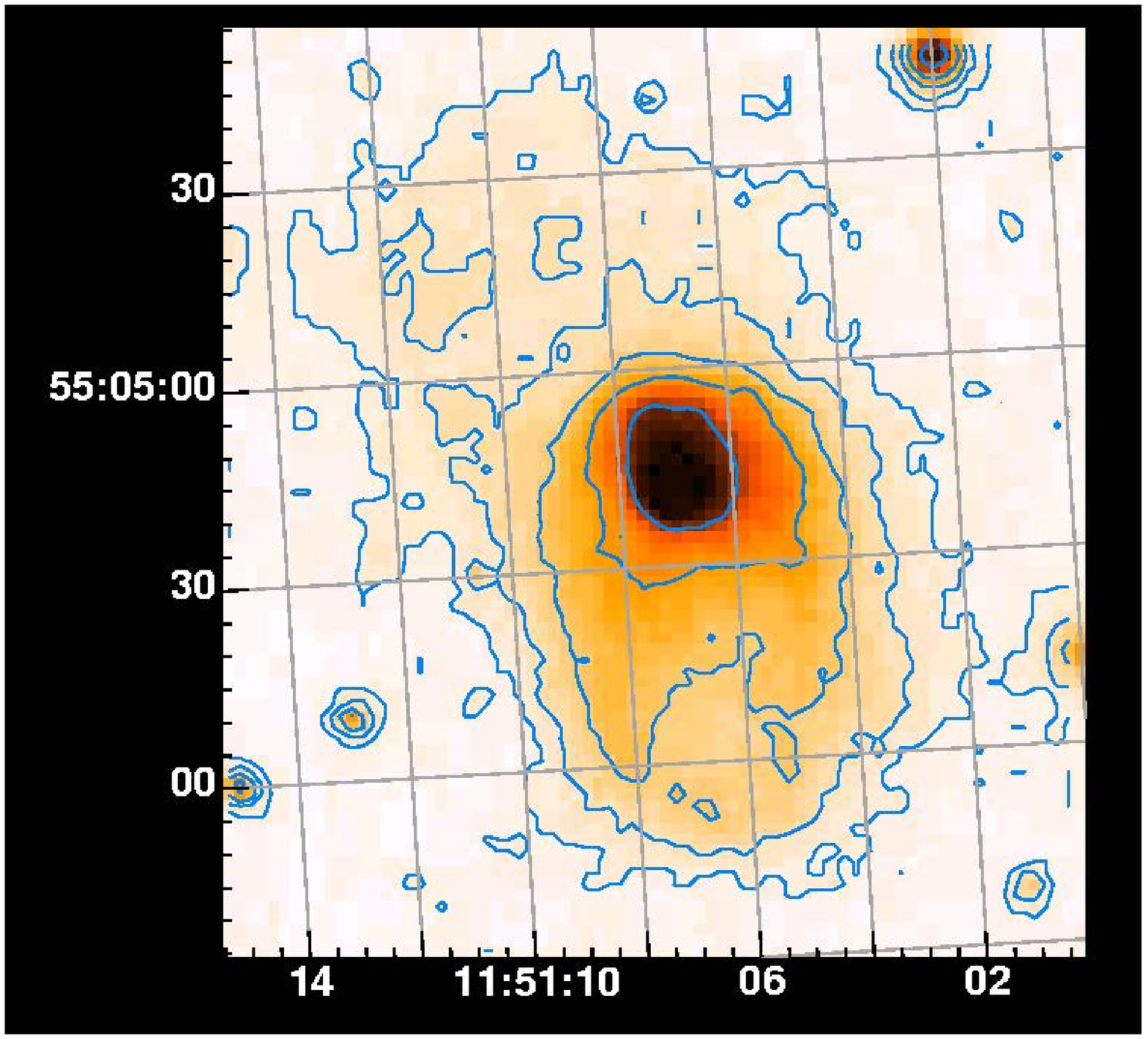,width=8.5cm,angle=-0,clip=}}

\vspace*{-6.5cm}
\raggedleft{\epsfig{file=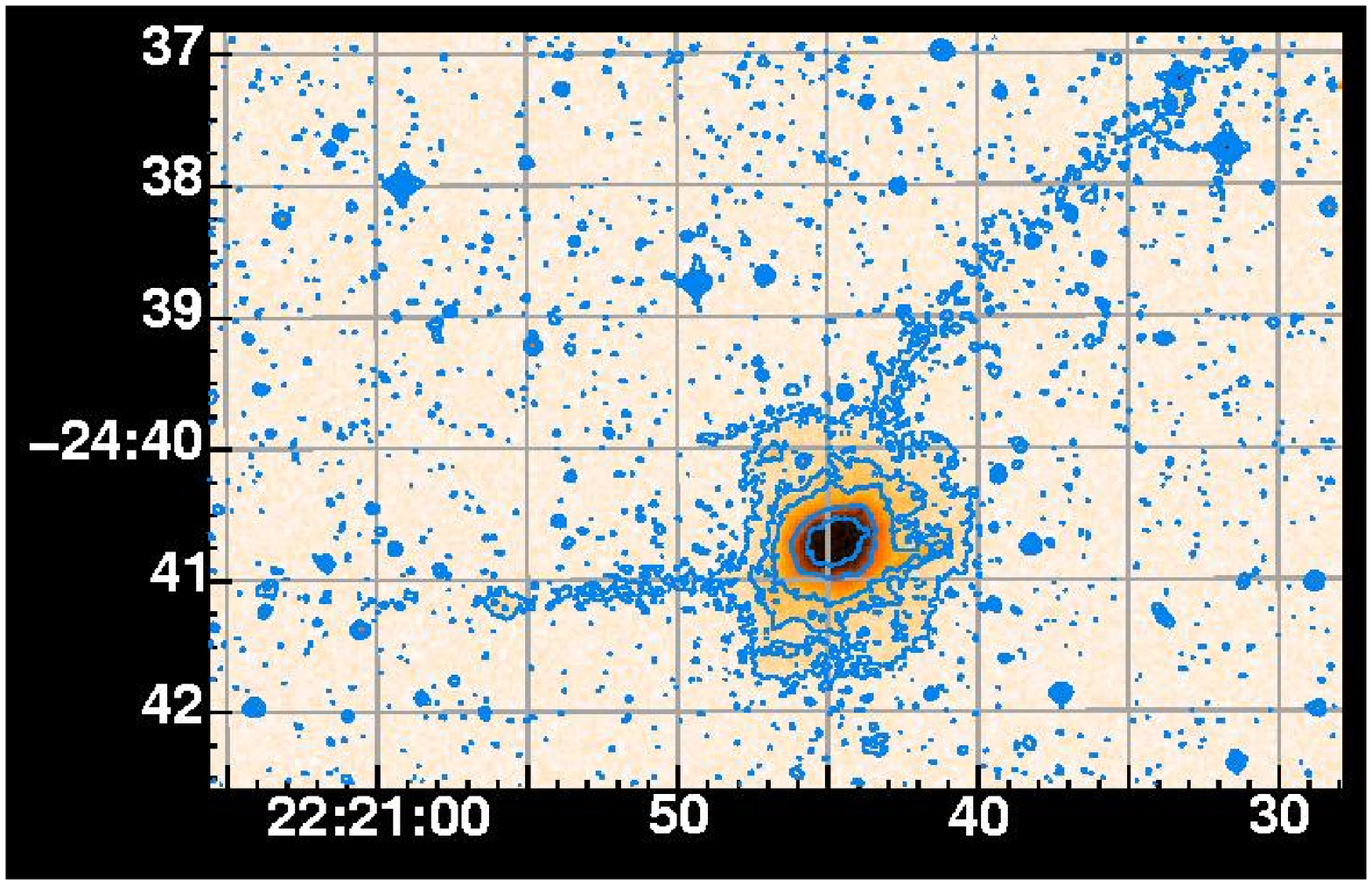,width=8.5cm,angle=-0,clip=}}

\vspace*{1.0cm}
\caption{DSS optical images of NGC 3921 (left, R$-$band) and NGC 7252 (right, B$_{j}-$band). The bright loop to the south and fainter tail to the northeast in NGC 3921, and the tails to the northwest and east in NGC 7252 are clearly visible. These loops and tails are typical of merger-remnant galaxies.}\label{opt}

\end{figure*}

The merger-remnant systems NGC 3921 and NGC 7252 are the
proto-typical products of major disk-disk merging, at the end of the merger 
evolutionary sequence (Read \& Ponman 1998). Their nuclei have
already coalesced, and their optical surface brightness profiles
are close to the r$^{1/4}$ de Vaucouleurs law seen in typical elliptical
galaxies (Schweizer 1996; 1982). However, they are not yet fully
relaxed, and still display the disturbed morphology of post-merger
objects, with tails and loops, which are the classic signature of
equal-mass mergers (Figure \ref{opt}).

These two objects are nearby, and have been well-studied across a
range of wavelengths. However it is only with the recent advent of
\xmm\ and \chandra\ that it has become possible to study their X-ray
emission in any detail. Our \xmm\ observations, and in the case of NGC
7252, archival \chandra\ data, in addition to the existing
multi-wavelength observations, therefore provide a new opportunity
to study the X-ray emission - from both extended and point-like
sources - arising in post-merger, proto-elliptical galaxies.

NGC 3921 and NGC 7252 have different physical properties (see Table
\ref{props}), most likely as a result of their different post-merger
ages, which are $\sim$0.7 and $\sim$1.0 Gyr respectively (Schweizer
1996; Hibbard \& Mihos 1995) and differences in their parent
galaxies. NGC 3921 is believed to be the product of a
gas-poor$-$gas-rich, Sa$-$Sc (or S0$-$Sc) merger (Hibbard \& van
Gorkom, 1996), whereas NGC 7252 is the product of gas-rich$-$gas-rich,
Sc$-$Sc merger (Hibbard et al. 1994). As NGC 7252 is more evolved than
NGC 3921, its optical emission is more relaxed and symmetrical. The
gas-rich nature of its progenitor galaxies has induced a vehement
starburst, with more luminous young globular clusters arising from
this starburst than are seen in the gas-poor$-$gas-rich merger remnant
NGC 3921 (see Table \ref{props}). These galaxies offer us an ideal
laboratory to investigate how the different merger histories and
merger ages of these galaxies are reflected in their X-ray
emission. Hence, we investigate both the extended hot diffuse gas
components and X-ray point source (XRP) populations in these galaxies.

\begin{table*}

\begin{center}

\begin{tabular}{lcc}

\hline      
  {                       }          &    {  NGC 3921   }               & {  NGC 7252 }                 \\
\hline      
Parent galaxies                      &      Sa$-$Sc (or S0$-$Sc)            &   Sc$-$Sc                       \\
                                     &         gas-poor + gas-rich      &   gas-rich + gas-rich         \\
Age / Gyr since peri-galacticon      &     $\sim$0.7$^{1}$               &  $\sim$1.0$^{2}$                  \\
Log (L$_{B}$ / \Lsolar)              &            10.76$^{3}$           &        10.81$^{4}$            \\
Log (L$_{FIR}$ / \Lsolar)            &            10.18$^{4}$           &        10.41$^{4}$            \\
S$_{60}$ / S$_{100}$                 &            0.52$^{5}$            &        0.57$^{5}$             \\
Optical isophotes of main body       &      `sloshing'                  &     symmetrical                 \\
Central structure                    & young stellar populations are    &  equilibrated central  disk of\\
                                     & asymmetrically distributed       & molecular and ionised gas     \\
 Nucleus                             &          LINER                   &     quiescent                 \\
 Young globular clusters             & $-9 < {\rm M_{\rm v}} < -13$    &  $-9 < {\rm M_{\rm v}} < -16$\\
Distance / Mpc                       &   77.9                          &  63.0                  \\
kpc / \arcsec                        & 0.360	 			&   0.304                       \\
Galactic line-of-sight $N_H$ / x10$^{20}$ cm$^{-2}$ & 1.14$^{6}$ & 1.95$^{6}$ \\
\hline
      
\end{tabular}
\caption{Summary comparison of the properties of NGC 3921 and NGC 7252 (for H$_{0} = 75$ km s$^{-1}$ Mpc$^{-1}$) $^{1}$Schweizer 1996; $^{2}$Hibbard \& Mihos 1995; $^{3}$Liu \& Kennicutt (1995); $^{4}$Fabbiano et al. (1992); $^{5}$Read \& Ponman (1998); $^{6}$Dickey \& Lockman (1990). }\label{props} 

\end{center}
\end{table*}

Ultra-luminous X-ray point sources (ULXs), with X-ray luminosities in
excess of $10^{39}$ erg s$^{-1}$, were detected in normal and
star-forming galaxies with the Einstein Observatory and ROSAT
(Fabbiano 1995, Roberts \& Warwick 2000), but it is only with the
advent of \xmm\ and \chandra\ that these studies could be extended
outside our local environment. The precise nature of these ULXs is
still open to debate. King (2002) suggests that, in addition to the
possible population of ULX supernovae remnants (SNR), there are in
fact two classes of ULX arising from super-Eddington mass inflow onto
compact objects: thermal-timescale mass transfer in high-mass X-ray
binaries (HMXBs), and long-lasting transient outbursts in low-mass
X-ray binaries (LMXBs). 

The donor stars of
HMXBs are expected to be O$-$B stars, with masses \gs 10 \Msolar, at
the end of their main sequence lifetime. This sets the timescale for
the evolution of HMXBs. The main sequence lifetimes of these donor
stars are short ($<$300 Myr), so HMXBs are associated with recent
star-formation. The ULXs observed in old stellar populations,
e.g. elliptical galaxies, must therefore be of the LMXB type
(e.g. Galactic micro-quasars), which is both longer-lived and takes
longer to evolve. Hence, it is HMXBs that we expect to dominate in the
merger-remnants NGC 3921 and NGC 7252, both of which have substantial
recent star-formation (Schweizer et al. 1996; Miller et al. 1997).

An alternative theory is that ULXs are fuelled by accretion onto
intermediate-mass black holes (IMBH, 10$^2-$10$^4$ \Msolar), formed
either from the collapse of massive population III stars or from
merging or accretion in dense stellar regions (e.g. Colbert \&
Mushotsky 1999; Ebisuzaki et al. 2001). This model is proposed to eliminate the necessity for super-Eddington accretion that HMXB models require.

Massive X-ray halos (up to 10$^{10}$\Msolar\ of hot gas) are evident
in elliptical galaxies (e.g. Forman et al. 1985; Fabbiano et
al. 1992). However, Read \& Ponman (1998) studied the X-ray evolution
of merging galaxies, with eight galaxies arranged in evolutionary
order, and found that late-stage merger remnants such as NGC 3921 and
NGC 7252 are under-luminous in X-rays when compared with typical
ellipticals. Log [(\lx/erg s$^{-1}$ / (L$_{B}$/\Lsolar)] $\sim$ 30.5
for a typical elliptical galaxy (from the Einstein observations of
Fabbiano et al. (1992)), yet our estimates of the luminosities of the
hot diffuse gas components give log [(\lx/erg s$^{-1}$ /
(L$_{B}$/\Lsolar)] $\sim$ 29.84 and 29.61 for NGC 3921 and NGC 7252
respectively, across the same energy range (0.2-3.5 keV) as the
Fabbiano et al. results. If, as is expected, these two galaxies are to
eventually evolve into normal ellipticals, their X-ray halos must be
rebuilt.

Three hypotheses have been proposed for the regeneration of the hot
halo: the infall of tidally stripped cold gas; mass loss from strong
stellar winds and the return of hot gas expelled from the main body of
the remnant during the nuclear merger stage. The sensitivity of
\xmm\ offers us the best opportunity to investigate these scenarios.

We have $\sim$25 ksec \xmm\ European Photon Imaging Camera
(EPIC) observations of NGC 3921 and NGC 7252, together with a 28.6
ksec archival \chandra\ observation of NGC 7252, with which we can
study both the ultra-luminous discrete source population and the hot
diffuse gas in these two galaxies. We adopt H$_{0} =$ 75 km s$^{-1}$
Mpc$^{-1}$ throughout this work.

The paper is structured as follows: \S2 describes the observations and
data reduction; in \S3 and 4, images and spectral fitting for NGC 3921
and NGC 7252 respectively are presented; in \S5, the ultra-luminous
discrete source population and the hot diffuse gas in the two galaxies
are discussed and compared. \S6 presents a summary of our results.

\begin{figure*}

\centerline{\epsfig{file=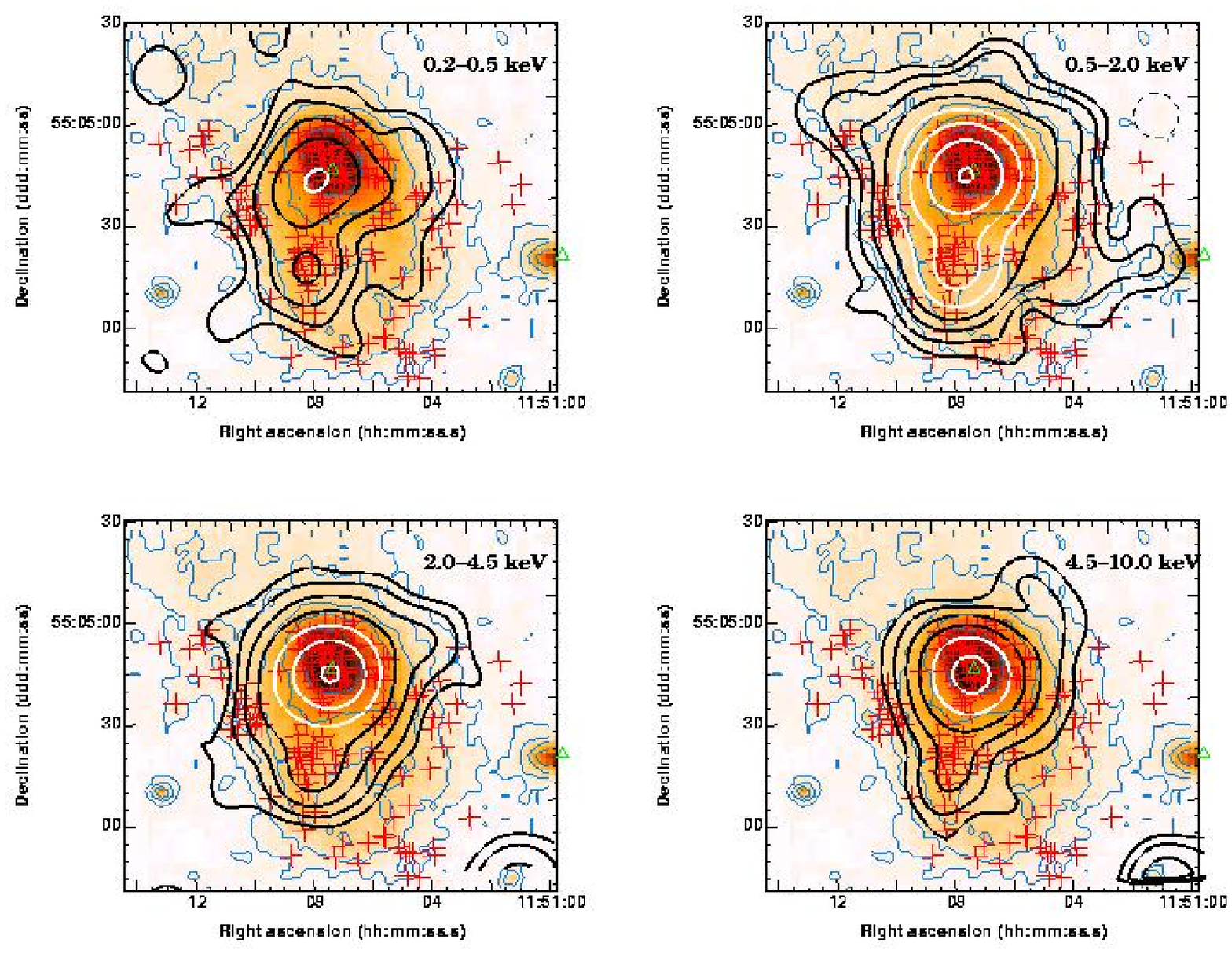,width=16cm,angle=-0,clip=}}

\caption{DSS optical (R$-$band) images of NGC 3921 (greyscale with thin grey contours) with all-EPIC X-ray emission contours (thick black and white, as described in $\S$ 2) overlaid. Top left: 0.2$-$0.5 keV; top right: 0.5$-$2.0 keV; bottom left: 2.0$-$4.5 keV; bottom right: 4.5$-$10.0 keV. The X-ray contour levels increase by factors of two from 1.375x10$^{-6}$ counts s$^{-1}$. Crosses: globular clusters and stellar associations (Schweizer et al. 1996); triangles: NGC 3921 and the dwarf galaxy 2MASX J11505929+5504133 to the east (Two Micron All-Sky Survey Team, 2003. Final release). The peak of the nuclear X-ray emission is off-set from the centre of the optical emission. The dashed circle in the top right plot represents the half-energy width (14\arcsec). The pixel sizes are 2.5\arcsec\ by 2.5\arcsec.}\label{3921ebands}
\end{figure*}

\section{Observations and data reduction}\label{obsanddr}

\subsection{\xmm\ observations: NGC 3921 and NGC 7252}

\begin{figure}
\vspace{-0.75cm}
\centerline{\epsfig{file=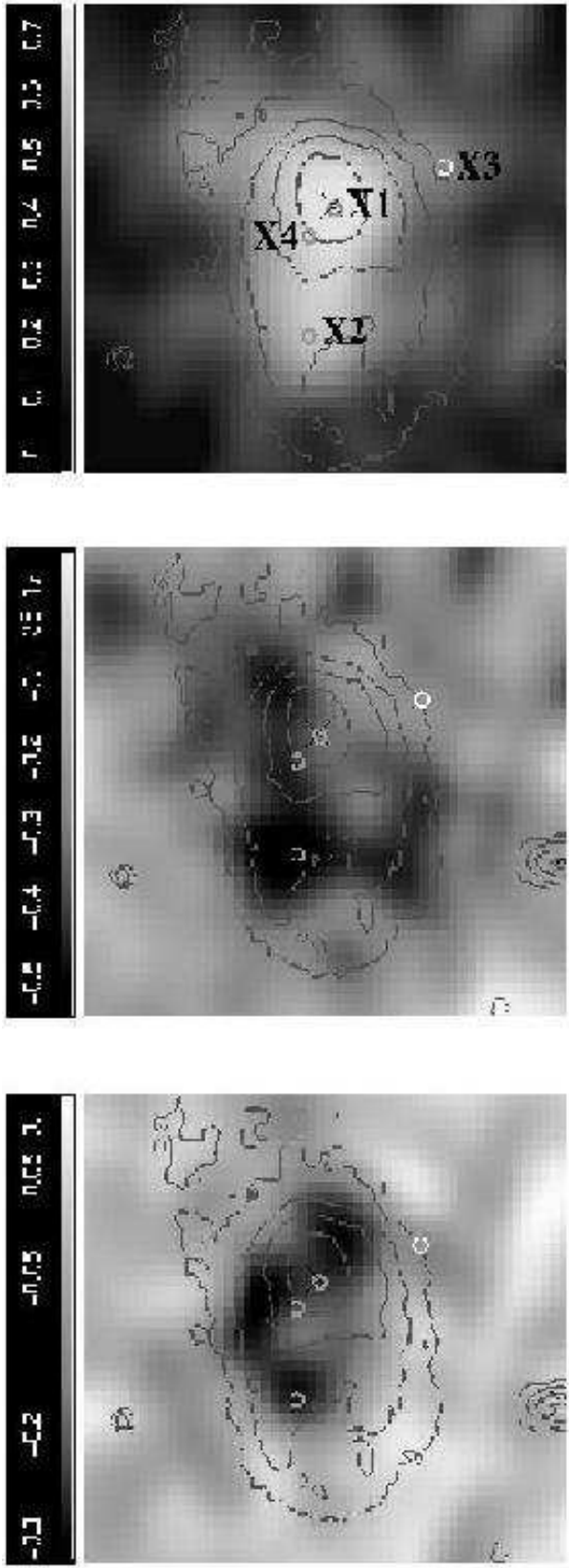,width=8.0cm,angle=-0,clip=}}
\caption{Hardness ratio maps for NGC 3921, with optical contours from the DSS R$-$band image overlaid. Top: HR $_{1}$(soft); middle HR$_{2}$ (medium) bottom: HR$_{3}$ (hard). See \S\ref{3921spattxt} for details. Lighter grey represents harder emission. The cross locates the co-ordinates of NGC 3921 (RA(J2000) 11h51m06.9s, Dec(J2000) +55d04m43s; Two Micron All-Sky Survey Team, 2003. Final release), and the circles the X-ray sources detected in the galaxy, which are labelled in the top plot as in Table \ref{3921srcs}.}\label{3921hr}
\end{figure}

\begin{figure}
\centerline{\epsfig{file=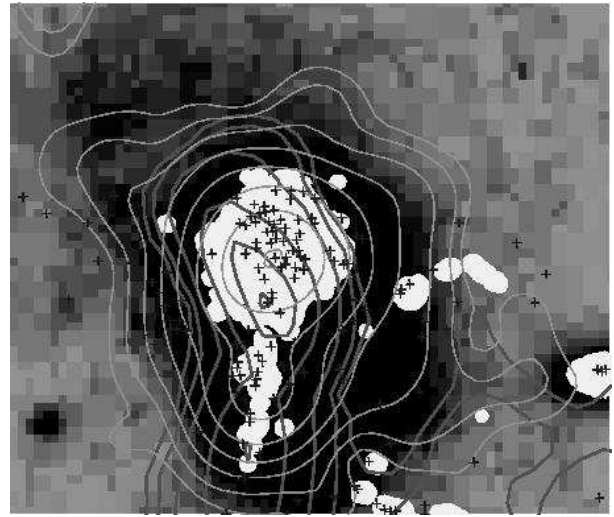,width=8.0cm,angle=-0,clip=}}\caption{Hot, warm and cold gas in NGC 3921. Light contours - all-EPIC X-ray emission, 0.5$-$2.0 keV, as described in $\S$ 2. White patches: H$\alpha$ (Hibbard \& van Gorkom 1996); grey scale: DSS R$-$band image; dark contours: HI (Hibbard \& van Gorkom 1996); crosses - globular clusters and stellar associations (Schweizer et al. 1996). 3921-X2 is located in the same region as the tails of H$\alpha$ and HI. }\label{3921Hregs}
\end{figure}

\begin{figure}
\centerline{\epsfig{file=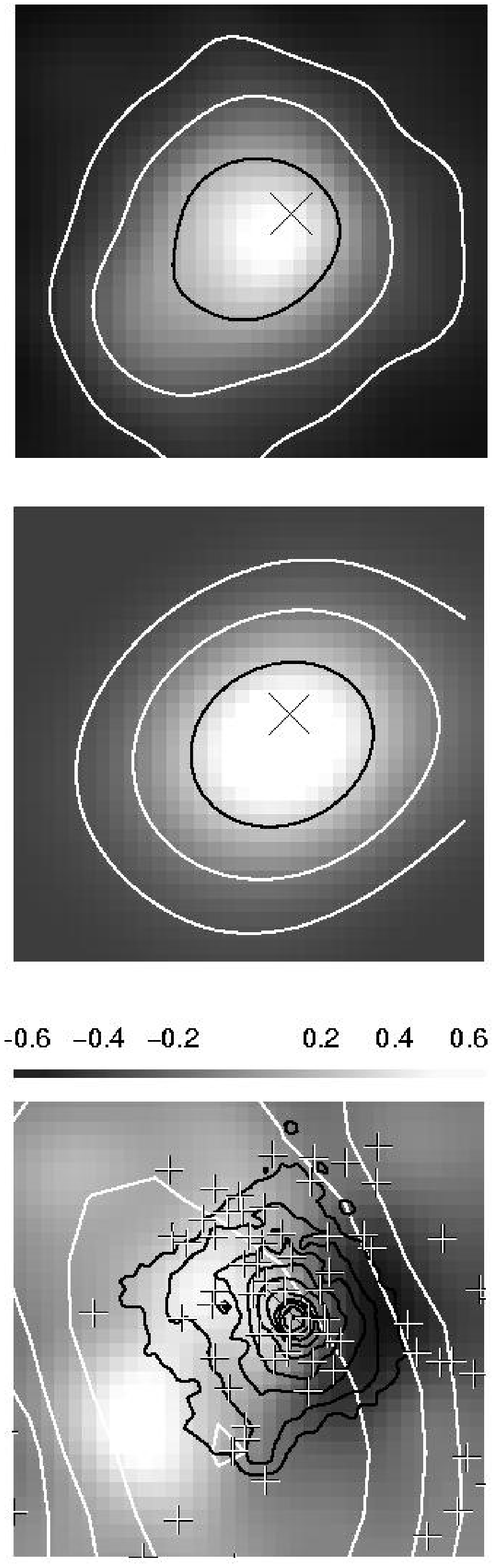,width=5.7cm,angle=-0,clip=}}
\caption{Spatial fitting to the 0.3$-$1.2 keV, MOS1, MOS2 and pn mosaiced and smoothed image (top) of the main body of NGC 3921. Top: mosaiced image, with contour levels overlaid; middle: best-fitting 2-D $\beta$-model, plus constant background, with the same contour levels as for the image overlaid; bottom: residuals ($data-model$) plus colour bar, with units in counts. The H$\alpha$ contours (black), HI contours (white) and candidate globular clusters and stellar associations (crosses; Schweizer et al. 1996) are superimposed upon the residuals. The diagonal cross marks the optical centre of NGC 3921. Each plot covers exactly the same region, of size 37.4\arcsec x37.4\arcsec, centred on the nucleus. The model is only fitted to the soft ($<$ 1.2 keV) emission in order to exclude contributions from the LINER whilst retaining sufficient emission (\gs 0.5 keV) which is not strongly absorbed. See \S\ref{3921spattxt} for details of the fitting.}\label{3921spat}
\end{figure}

The details of the \xmm\ observations of NGC 3921 and NGC 7252 are
listed in Table \ref{obs}. Only the EPIC data for these two galaxies
are discussed in this paper; there are too few counts for meaningful
analysis of the Reflection Grating Spectrometer data. The \xmm\
Science Analysis System (SAS), version 5.4.1, was used to process the
observation data files into calibrated events lists, clean up
background flaring (with flag $=$ $\#$XMMEA$\_$EM for the MOS data,
and $\#$XMMEA$\_$EP for the pn data) and filter the events. For the
MOS1 and MOS2 data, single, double and quadruple pixel events are
selected, and for the pn data single and double events. The `good
times' remaining for each camera after the cleaning process are listed
in Table \ref{obs}. For both galaxies, the data were very clean, with
only a small fraction requiring removal because of background flaring.

SAS and HEASARC's FTOOLS version 5.2 were used to create
background-subtracted, exposure-corrected, all-EPIC mosaiced images in
four energy bands: 0.2$-$0.5, 0.5$-$2.0, 2.0$-$4.5 and 4.5$-$10.0
keV. The exposure maps were created using {\it eexpmap}, and the
background maps with {\it esplinemap}, following the prescription
given in the SAS data analysis threads. The mosaiced images were
simply-smoothed with a Gaussian of FWHM 6.25\arcsec\ (compared with
\xmm\ spatial resolution $\sim$7\arcsec) to produce the X-ray contour
maps presented in this paper. Source-searching was carried out for
each of the three EPIC instruments in all four energy bands
simultaneously, using the SAS source-searching packages.

Two-dimensional surface-brightness fitting was carried out on the 0.3$-$1.2 keV mosaiced raw images, using the {\it Sherpa} software from the {\it Chandra Interactive Analysis of Observations} (CIAO) software package, version 2.3. Exposure correction is carried out on the model rather than the data, so as not to interfere with the fitting statistics. The background is modelled with a constant amplitude component, allowed as a free parameter in the fitting process. The area over which the fitting is carried out is small (37.4\arcsec x37.4\arcsec\ and 58.3\arcsec x40.7\arcsec\ for NGC 3921 and NGC 7252 respectively, centred on the nucleus in both cases), so exposure and PSF effects on the background are negligible..  The image is convolved with the combined all-EPIC point spread function, weighted to account for the greater effective area of the pn camera compared with the MOS cameras. The best-fitting model is found using the Cash statistic.

The hardness ratio maps of Figures \ref{3921hr} and \ref{7252hr} were calculated in the following way. First, the (unsmoothed) 0.2$-$0.5, 0.5$-$2.0, 2.0$-$4.5 and 4.5$-$10.0 keV background-subtracted, exposure-corrected, all-EPIC mosaiced images described above had pixels with $<$ 1x10$^{-6}$ counts per second set to zero. The hardness ratios were then calculated using  {\rm $HR_{n} = (E_{n+1} - E_{n}) / (E_{n+1} + E_{n})$ }, where n $= 1 - 3$, E$_{1}$ is the 0.2$-$0.5 keV image, E$_{2}$ is the 0.5$-$2.0 keV image, E$_{3}$ is the 2.0$-$4.5 keV image and E$_{4}$ is the 4.5$-$10.0 keV image. The hardness ratio maps were then simply-smoothed, with a Gaussian of FWHM 6.25\arcsec. Pixels with values $\pm$1 in these maps were then set to zero, to remove areas where photons are only detected in one band. For NGC 7252, there is no HR$_{3}$ map, as there is so little emission in the hardest energy band. Within the outermost optical contours plotted in Figures \ref{3921hr} and \ref{7252hr}, the uncertainties in the hardness ratios are of the order a few percent for HR$_{1}$ and HR$_{2}$, and $\sim$10\% for NGC 3921 HR$_{3}$.

\subsection{\chandra\ observation: NGC 7252}

NGC~7252 was observed with the Chandra Observatory (observation ID
2980) on the 17th November 2001 for approximately 28.6 ks, with the
back-illuminated ACIS-S3 CCD chip at the focus. Data products,
correcting for the motion of the spacecraft and applying instrument
calibrations, were produced using the Standard Data Processing system
at the Chandra X-ray Center (CXC). A preliminary analysis of these
products was then performed using the CXC CIAO software suite (version
2.2.1).

A light-curve extracted from a large non-source area revealed a
moderate flare, with a sharp peak at $\sim$twice the quiescent
background level, occurring over the latter quarter of the
observation. Therefore, we performed our analysis on the
flare-subtracted data (which amounted to 22.15 ks). 

\begin{table*}

\begin{center}
\begin{tabular}{cccccccc}

\hline
object   & date & camera & obs time /ksec & filter & window & good time / ksec \\
\hline
NGC 3921 & 27/04/2002    & MOS1 & 26.0 & medium & PrimeFullWindow & 24.8  \\
         &               & MOS2 & 26.0 & medium & PrimeFullWindow & 25.1  \\ 
         &               & pn   & 23.7 & medium & PrimeFullWindow & 20.5  \\ 
\hline 
NGC 7252 & 13-14/11/2001 & MOS1 & 27.6 & medium & PrimeFullWindow & 26.0  \\ 
         &               & MOS2 & 27.6 & medium & PrimeFullWindow & 26.2  \\ 
         &               & pn   & 25.0 & medium & PrimeFullWindow & 20.3  \\   
\hline

\end{tabular}

\caption{Details of \xmm\ observations of NGC 3921 and NGC 7252. }\label{obs} 

\end{center}
\end{table*}

\section{NGC 3921}

\subsection{Spatial data}\label{3921spattxt}

The background-subtracted, exposure-corrected, smoothed X-ray contour maps of NGC 3921, prepared as described in \S2.1, are presented in Figure \ref{3921ebands}, superposed on the Digitized Sky Survey (DSS) optical (R$-$band) image. Four energy bands are shown: 0.2$-$0.5, 0.5$-$2.0, 2.0$-$4.5 and 4.5$-$10.0 keV. The strongest emission is seen in the 0.5$-$2.0 keV range, but even at high energies (4.5$-$10.0 keV), there is still significant emission, presumably, at least in part, from the low-ionization nuclear emission-line region (LINER) residing in this galaxy (Stauffer 1982a,b). The peak of the X-ray emission is offset to the south east of the optical peak (central triangle), with the offset being greatest in the softest band. This is indicative of a possible absorption effect, obscuring the true X-ray nucleus in the soft band, which is consistent with the results of model fitting to the nuclear spectra discussed in \S3.2. The X-ray emission is not extended beyond the main body of the galaxy.

\begin{table*}

\begin{center}
\begin{tabular}{ccccc}

\hline
source  & RA(J2000) / h:m:s & Dec(J2000) / d:m:s & detection likelihood & 10$^{-2}$ counts s$^{-1}$ \\
\hline
3921-X1 & 11:51:06.96 & 55:04:42 & 167.5  & 8.913$\pm$0.272 \\
3921-X2 & 11:51:07.92 & 55:04:13 &  56.7  & 2.652$\pm$0.134 \\
3921-X3 & 11:51:03.84 & 55:04:50 &  74.2  & 0.365$\pm$0.081 \\
3921-X4 & 11:51:07.68 & 55:04:35 &  56.5  & 1.723$\pm$0.312 \\
\hline

\end{tabular}

\caption{Sources detected in NGC 3921, using the SAS source-searching software. The positions of the sources are plotted in Figure \ref{3921hr}.}\label{3921srcs} 

\end{center}
\end{table*}

Figure \ref{3921hr} shows the three hardness ratio (HR) maps calculated for NGC 3921 as described in $\S$\ref{obsanddr}, with optical contours from the DSS (R$-$band) image superposed. In these plots, darker greys represent softer emission, and lighter shades are harder. The detected sources are marked with circles, and identified according to their listing in Table \ref{3921srcs}. Most emission is in the 0.5$-$2.0 keV range; the HR$_{1}$ map shows the hardness peaking (from $\sim$0.35 in the outer regions to $\sim$0.74) in the centre of the galaxy; for HR$_{2}$ and HR$_{3}$, the hardness ratio remains negative, but broadly declines in the inner regions by a factor $\sim$2.

These plots reveal an irregular, energy-dependent structure to the X-ray emission. Although the central emission in the softest bands (HR$_{1}$, top) is relatively symmetrical about the centre of the galaxy, at higher energies (HR$_{2}$ and HR$_{3}$), this is certainly not the case. The hardness peak in HR$_{2}$ near the centre of the optical emission probably results from the LINER emission, which peaks in X-rays in the 2.0$-$4.5 keV band (see the spectrum in Figure \ref{3921-centre-spec}). However, outside this very central region, the asymmetry may be due to irregular dust obscuration, unresolved X-ray point sources, or the disturbed distribution of the hot diffuse gas components. As is confirmed by the spectra plotted in Figures \ref{3921-centre-spec}, \ref{3921-blob-spec} and \ref{3921-s3-spec}, the LINER has much harder X-ray emission than the discrete sources, which have little emission above 5 keV.

Interestingly, in Figure \ref{3921ebands}, we see what appears to be a `bridge' of X-ray emission, in the 0.5$-$2.0 keV band, from the main body of the remnant towards the dwarf galaxy 2MASX J11505929+5504133. Schweizer (1996) finds a recession velocity for this galaxy (S96 G4 in his notation) of 5712 km s$^{-1}$, so close to that of NGC 3921 (5838 km s$^{-1}$), that it is likely to be physically associated with it.  The existence of the bridge is suggestive of the formation of the dwarf galaxy from tidal debris ejected during the merging of the parent galaxies, as may be the case at the end of the southern tail of the Antennae galaxies (e.g. Zwicky 1956; Deeg et al. 1998; Hibbard et al. 2001). However, the optical surface brightness of S96 G4 is much higher than that seen in any such candidate dwarf galaxies (Schweizer 1996). The 0.5$-$2 keV emission in the bridge region is $\sim 4\sigma$ above the background emission, and so this feature is likely to be real. In addition to the suggested 'bridge', there are two arms of H$\alpha$ emission from the main body of NGC 3921 towards this galaxy (Hibbard \& van Gorkom 1996), which  correspond closely to groups of young stellar
associations and candidate young clusters imaged with the Hubble Space Telescope (HST; Schweizer et al. 1996). These are visible in Figure \ref{3921Hregs}, above and below an imaginary line connecting the dwarf galaxy with the nucleus of NGC 3921 (the lower arm is right at the bottom edge of the frame). We tentatively suggest the possibility that these two `ridges' of H$\alpha$ emission, which are nearly orthogonal to the main tidal-tail structures, are wakes, produced by this small, dense disk galaxy having punched through some of the dense HI gas south of NGC 3921 at relatively high velocity (several hundred km s$^{-1}$). 

Source-searching with the SAS software finds four sources in NGC 3921, although there are too few counts from any of the sources to fit the spatial extent. The minimum detection likelihood for inclusion of the source in the output list is set at 8, so that the probability that a random Poissonian fluctuation has caused the observed source counts is 3.35x10$^{-4}$. 3921-X1, the most significant source, is coincident with the nucleus of NGC 3921 (RA(J2000) 11h51m06.9s, Dec(J2000) +55d04m43s; Two Micron All-Sky Survey Team, 2003. Final release), where the LINER resides. There is also a bright source $\sim$ 0.5\amin\ south of the nucleus (3921-X2), associated with the southern tail and a concentration of stellar associations and/or globular clusters. This source is visible in all the energy bands, and is softer than the central region (which contains the LINER). There is some harder emission to the south of the main peak of 3921-X2, observable in the HR$_{3}$ map. 3921-X3, to the north east, is detectable as a slight bulge in the X-ray contours in Figure \ref{3921ebands}, and is visible in the hardness ration maps. 

Figure \ref{3921Hregs} shows the H$\alpha +$[NII] (KPNO 2.1m
telescope) and HI (VLA) emission detected in NGC 3921 (Hibbard \& van
Gorkom 1996). There is a ridge of H$\alpha$ emission, which is
associated with recent star formation, along the southern tidal
tail. This ridge coincides with an extended ridge of HI in the same
direction. 3921-X2 appears to be associated with this region of star
formation and atomic gas in the southern tail, and clearly resides
within this tidal tail.

We model the surface brightness of the 0.3$-$1.2 keV, all-EPIC mosaiced image of the main body of NGC 3921, within a square of side 37.4\arcsec\ centred on the nucleus. The model is a $\beta$-model, with an adjustable power-law, of the form:

\begin{equation}
 f(r) = A[1 + (r/r_{0})^{2}]^{-\alpha}, 
\end{equation}

where $r_{0}$ is the core radius, $A$ is the amplitude at $r=0$, and $\alpha = 3\beta - \frac{1}{2}$. A constant background component is also included, and the model is convolved with the mosaiced point spread functions (PSF) of the three EPIC instruments. The energy is cut off at 1.2 keV in order to exclude the bulk of emission from the LINER, whilst extending sufficiently far beyond the energy range most affected by absorption (i.e. \gs 0.5 keV) to retain sufficient counts to constrain the fitted parameters.

Table \ref{3921spatfittab} lists the best-fitting parameters of this model for the main body of NGC 3921. Figure \ref{3921spat} presents the data, best-fitting model and residuals of the fit ($data-model$). The HI and H$\alpha$ contours are superimposed on the plot of the residuals. There is an excess to the south east of the nucleus, which is not spatially coincident with any globular cluster candidate. This is plausibly emission from 3921-X4, which we know has soft ($<$ 1.2 keV) emission (see \S\ref{3921spectxt}). However, we have too few counts from this source to allow us to successfully disentangle it from the hot diffuse gas emission and investigate its spectrum.

The power-law index, $\alpha$ is 2.93$\pm$0.08, and the core radius is 4.43$\pm$0.09 kpc (H$_{0} =$ 75 km s$^{-1}$Mpc$^{-1}$). The parameters of the spatial model should be regarded as simply describing the shape of the surface brightness distribution, rather than offering a physical interpretation of that shape. 

\begin{figure}
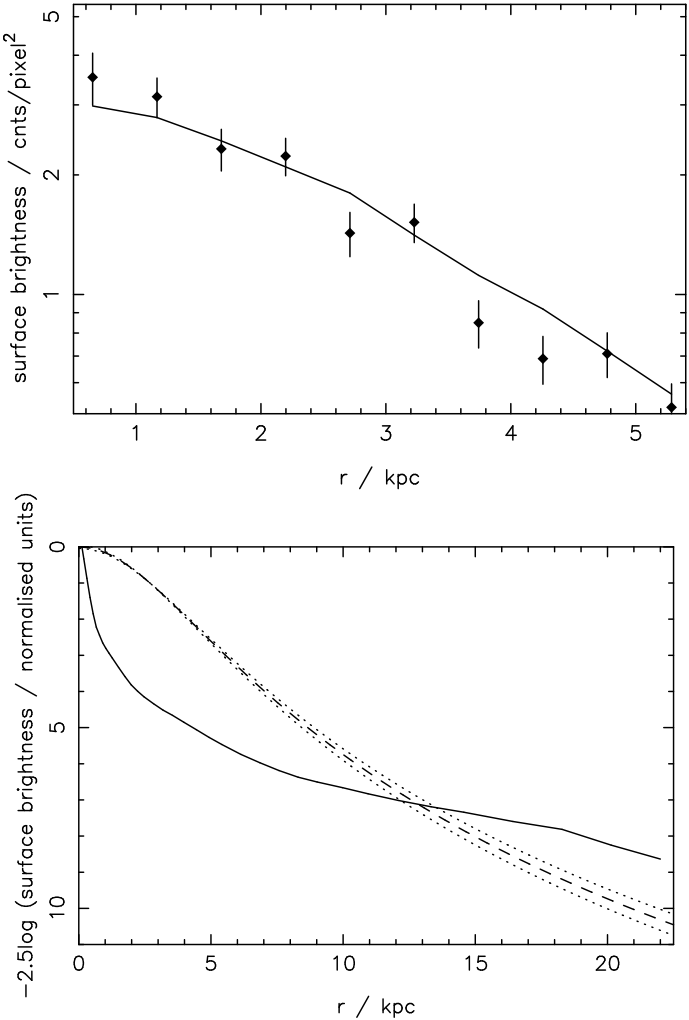

\centerline{\epsfig{file=3921-Xsbr.ps,width=6.5cm,angle=-90,clip=}}
\centerline{\epsfig{file=3921-sbr.ps,width=7.0cm,angle=-90,clip=}}
\caption{Top: diamonds: radial profile (raw data, binned over circular annuli) of the diffuse gas in NGC 3921, plus error bars, over the range of the 2-dimensional fitting; solid line: model radial profile, using the best-fitting 2-dimensional model parameters, convolved with the PSF. Bottom: dashed line: model radial profile of the hot diffuse gas in NGC 3921, using the fitted parameters from the best-fitting 2-D spatial fitting (see Table \ref{3921spatfittab}and \S\ref{3921spattxt}). The dotted lines represent the uncertainties in the model fitting. Solid line: mean V-band surface brightness profile (Schweizer 1996). The surface brightness has been normalised to unity at r$=$0 in both cases. }\label{3921sbr}
\end{figure}

Figure \ref{3921sbr} compares the modelled X-ray surface brightness distribution with the mean V-band radial profile. The hot diffuse gas emission has a very different profile from the optical profile, with a more flattened core, and a fairly uniform decline from \gs3 kpc. The raw X-ray data, together with the best-fitting model surface brightness distribution, are also shown in Figure \ref{3921sbr}.

\begin{table}

\begin{center}
\begin{tabular}{cccc}

\hline
object   & core radius / kpc & $\alpha$  & amplitude / cnts \\\hline
NGC 3921 & 4.43$\pm$0.09 & 2.93$\pm$0.08 & 5.42$\pm$0.20 \\
\hline
\end{tabular}

\caption{The results of 2-D spatial fitting to the 0.3$-$1.2 keV, MOS1, MOS2 and pn mosaiced images of the main body of NGC 3921. The fitted model is a $\beta$-model, as described in \S\ref{3921spattxt}. $r_{0}$ is calculated for H$_{0} =$ 75 km s$^{-1}$Mpc$^{-1}$.  Figure \ref{3921spat} presents the data, best-fitting models and residuals of the fits.}\label{3921spatfittab}
\end{center}
\end{table}

\subsection{Spectral fitting}\label{3921spectxt}

For the extraction of the spectra, source and background regions were
defined with the help of the DS9 display tool. The source regions are
shown in Figure \ref{3921-centre-spec}, and have radii 19.6\arcsec\
(nucleus), and 12.0\arcsec\ (3921-X2 and 3921-X3). The background
regions were defined within large circles on the same CCDs as the
sources, in regions uncontaminated by detected sources. There is
probably some contamination from nuclear emission in the off-nuclear
sources, but, from Figure \ref{3921sbr}, the surface brightness of the
nuclear region has dropped to $<$10\% of the peak value for the hot
diffuse gas at the distances of 3921-X2 and X3, so the X-ray emission
will be dominated by the off-nuclear sources. SAS tasks were used for
the extraction, and to generate the response products. The spectra
were grouped so that there were at least 20 counts per bin. Model
spectra were fitted via \xs\ minimisation using HEASARC's XSPEC
software package, version 11.2. All the models fitted include a
component for line-of-sight Galactic absorption, which is frozen in
the fitting process, as well as intrinsic hydrogen absorption, which
is allowed to vary. It is the fitted intrinsic absorption which is
quoted in Tables \ref{3921-centre-tab}, \ref{3921-src-tab},
\ref{7252-centre-tab} and \ref{7252-blob-tab}. The Galactic
line-of-sight equivalent hydrogen column density is given in table
\ref{props}.

\subsubsection{The nucleus}

\begin{figure}
\centerline{\epsfig{file=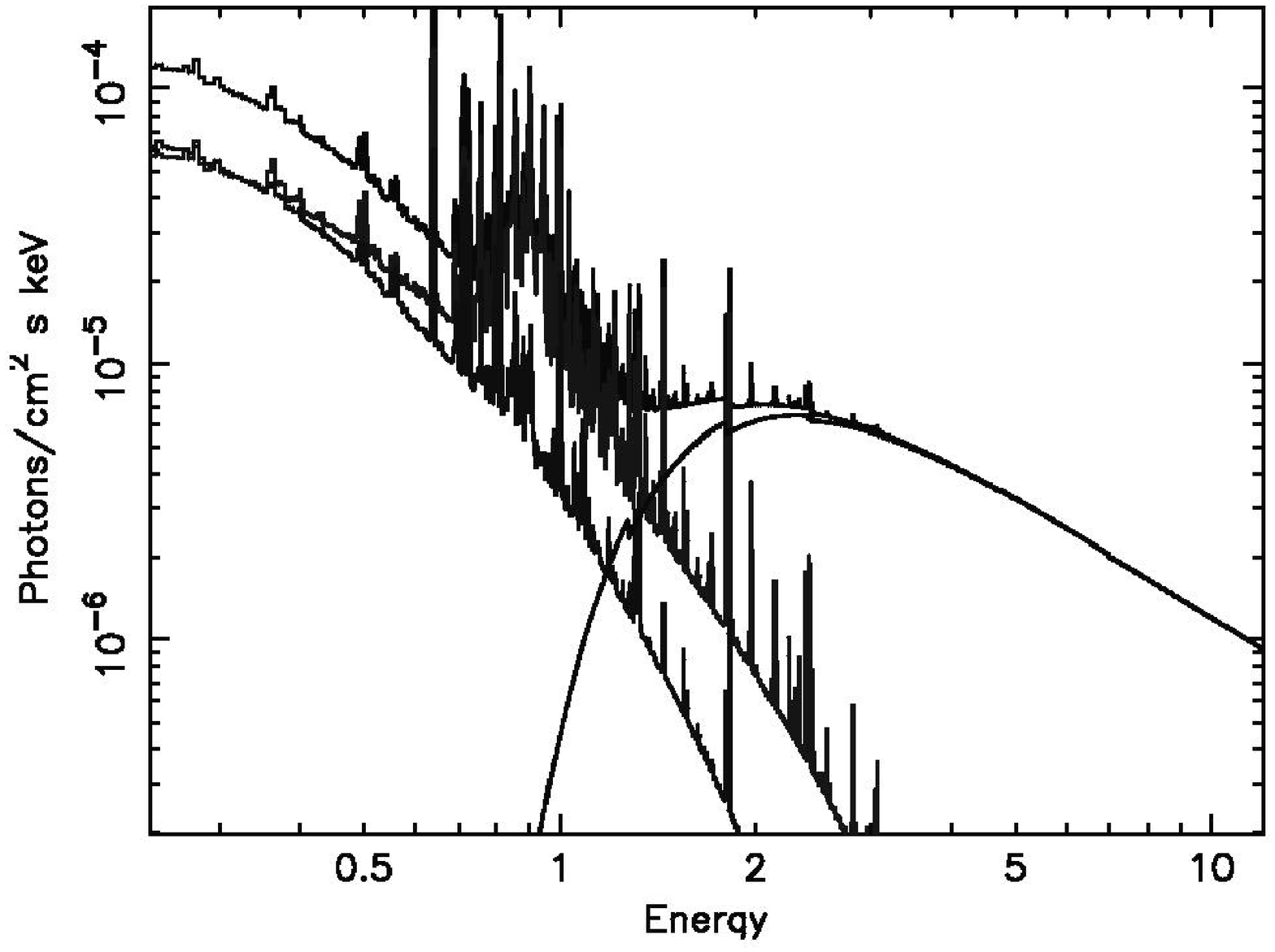,width=7.0cm,angle=-90,clip=}}
\centerline{\epsfig{file=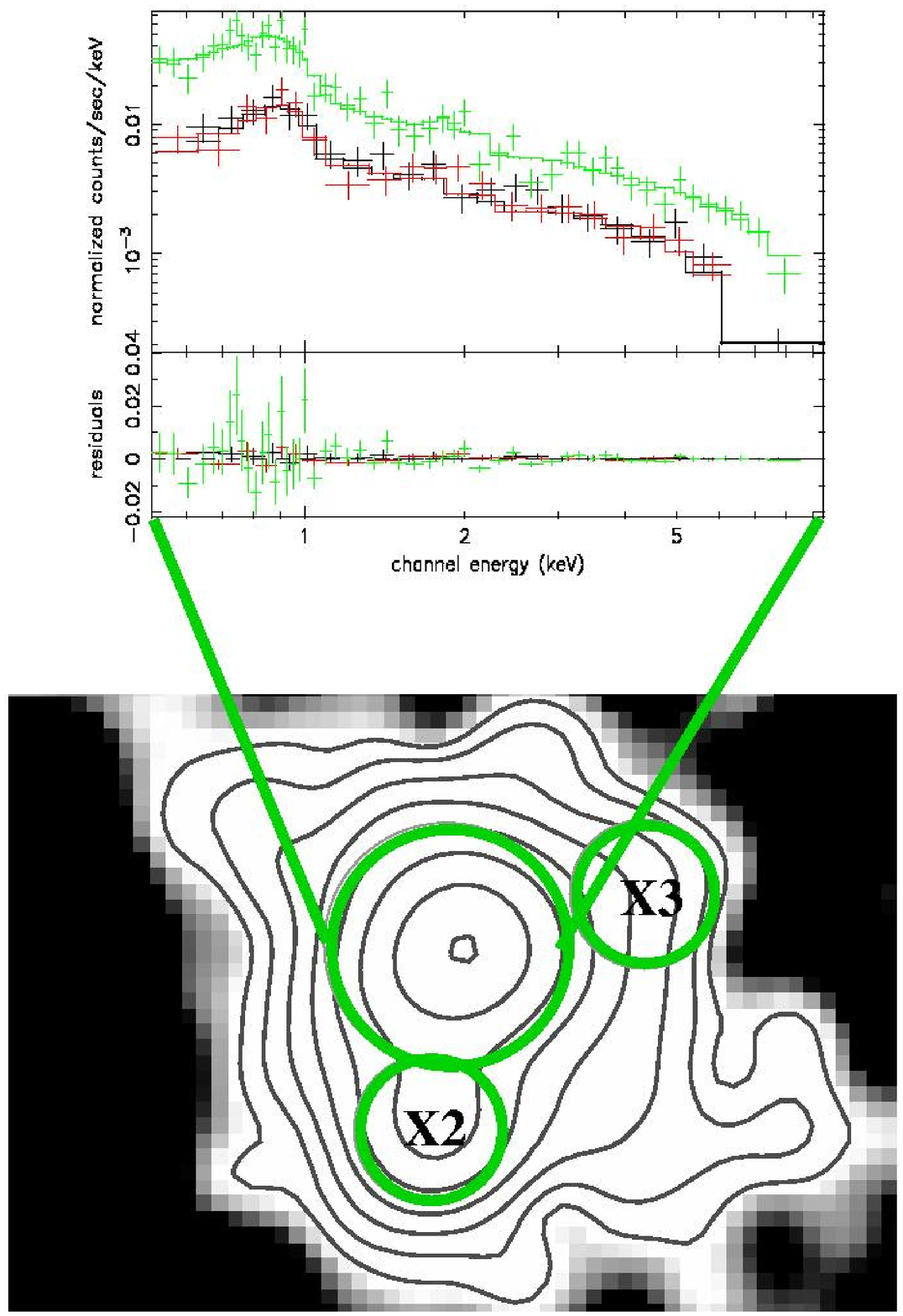,width=8.5cm,angle=-0,clip=}}
\caption{Spectral fit to the nucleus of NGC 3291. Top: spectral plot of this best-fitting model (top line) showing the two MeKaL components, which dominate at lower energies, and the substantial absorbed power-law component, which dominates at energies \gs\ 2 keV. Middle: crosses: pn (upper) and MOS (lower) spectra; solid lines: best-fitting model spectra. Bottom: X-ray contour map (all-EPIC mosaic, 0.5$-$2.0 keV) showing the central region from which the spectra were extracted. See Table \ref{3921-centre-tab} for details of the models, and \S\ref{3921spectxt} for description of the fitting and discussion.}\label{3921-centre-spec}
\end{figure}

\begin{table*}
\begin{center}

\begin{tabular}{lllll}

\hline
 {model}         & {MeKaL 1} & {MeKaL 2}     & {power-law} & {total} \\
\hline
 {kT / keV}        & {  0.79\ (0.73$-$0.85)} & {  0.45\ (0.37$-$0.55)}  & {  }    & {  }          \\
 {abundance }      & {  0.19\ (0.16$-$0.22)} & {  0.03\ (0.01$-$0.05)}  & {  }    & {  }          \\
 {photon index}    & {  }                       & {  }      & {  1.50\ (1.45$-$1.55)} & {  }       \\  
 {intrinsic $N_H$ / 10$^{22}$ cm$^{-2}$}    & {  }      & {  }    & {  1.93\ (1.73$-$2.14)} & {  }      \\ 
 {log (L$_{X}$ / erg s$^{-1}$)} & {  40.29\ (40.27$-$40.37)} & {  39.90\ (39.65$-$40.05)}  & {  41.33\ (41.32$-$41.37)} & {  41.38\ (41.36$-$41.43) }   \\
 {\xs  / d.o.f}      & {  }  & {  }      & {  }    & {  93.2 / 86} \\
\hline

\end{tabular}

\caption{Results from simultaneous fitting to the MOS1, MOS2 and pn, 0.5$-$10.0 keV, spectra of the nucleus of NGC 3921 (see Figure \ref{3921-centre-spec}). The model fitted has components for the Galactic extinction, two MeKaL components, and an absorbed power-law. Abundance is measured relative to solar. The luminosities quoted are unabsorbed, for H$_{0} = 75$ km s$^{-1}$ Mpc$^{-1}$. The errors quoted are the 90\% confidence limits on each single parameter of interest, and the hydrogen column density refers to the intrinsic absorption, which is fitted in addition to the fixed line-of-sight Galactic absorption.}\label{3921-centre-tab} 
\end{center}
\end{table*}

The 0.5$-$10.0 keV MOS1, MOS2 and pn spectra of the central region of NGC 3921, totalling 1987 counts, are shown in Figure \ref{3921-centre-spec}. The extraction  region is shown in the same figure. The spectra were modelled simultaneously by a two-component MeKaL emission spectrum for hot diffuse gas (Mewe et al. 1985; Mewe et al. 1986; Kaastra 1992; Liedahl et al. 1995) plus an absorbed power-law component. In Table \ref{3921-centre-tab}, the parameters of the best-fitting model are listed, together with a spectral plot of the various contributions from the model components. Two MeKaL components in the model results in a much better fit than a single component, as one might expect in this disturbed starburst system. 

The MeKaL components (kT $=$ 0.79 and 0.45 keV) contribute most of the energy between 0.5 and 2.0 keV. The harder end of the spectrum (\gs\ 3 keV) is well-modelled by an absorbed power-law component, with a photon index of 1.50, and intrinsic hydrogen column density, n$_{H} = 1.93{\rm x}10^{22}$ cm$^{-2}$. The dominant contributor to this luminous (log (L$_{X}$ / erg s$^{-1}) =$ 41.33) power-law component is probably the low-ionization nuclear emission region (LINER) residing in this galaxy. However, it is likely that there are also some unresolved X-ray point sources which contribute to the hard end of the spectrum. The overall quality of the fit is acceptable.

\subsubsection{3921-X2: the southern source}

\begin{figure}
\centerline{\epsfig{file=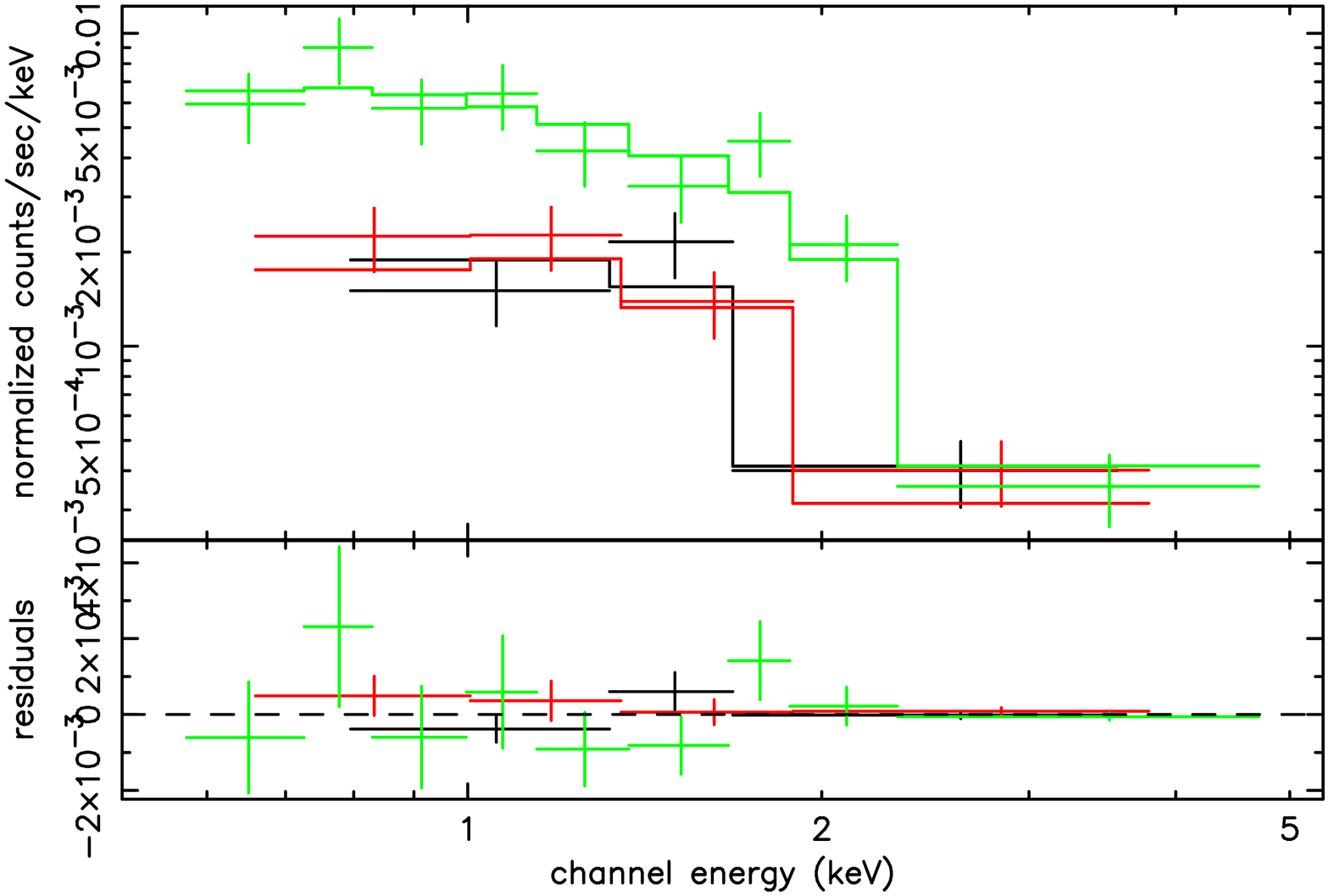,width=6.0cm,angle=-90,clip=}}
\caption{Spectral fit to 3921-X2, the southern source in NGC 3291. Crosses: pn (upper) and MOS (lower) spectra; solid lines: best-fitting disk blackbody model spectra. See Table \ref{3921-src-tab} for details of the models and \S\ref{3921spectxt} for description of the fitting and discussion. }\label{3921-blob-spec}
\end{figure}

\begin{table*}
\begin{center}
\begin{tabular}{lll}

\hline
{SOURCE}                       &  {\bf 3921-X2}                & {\bf 3921-X3}                   \\
{counts}                       &  {322}                        & {149}                           \\   
\hline
{POWER-LAW}                    &                               &                                 \\
{photon index}                 &{2.37\ (2.20$-$2.58)}       &{\bf  1.28\ (0.90$-$1.46)}     \\
{intrinsic $N_H$ / 10$^{22}$ cm$^{-2}$} &{0.21\ (0.16$-$0.26)} &{\bf  0.00\ ($<$0.13)}     \\
{log (L$_{X}$ / erg s$^{-1}$)} &{40.35\ (40.26$-$40.43)}    &{\bf  40.02\ (39.92$-$40.22)}  \\
{\xs / d.o.f}                  &{13.7 / 13}                    &                                 \\       
{CASH statistic / no. of bins} &                               &{\bf  120.6 / 133}                 \\
\hline
{MEKAL}                        &                               &                                 \\
{kT / keV}                     &{2.18\ (1.94$-$3.90)}       &{  63.3(13.3$->$9.9)}         \\
{abundance }                   &{0.00\ ($<$0.35)}       &{0.00\ ($<$546.3)}        \\
{log (L$_{X}$ / erg s$^{-1}$)} &{40.23\ (40.20$-$40.46)}    &{40.01\ (38.99$->$42.07)}     \\
{\xs / d.o.f}                  &{12.4 / 12}                    &                                 \\       
{CASH statistic / no. of bins} &                               &{  120.5 / 133  }                 \\    
\hline
{DISK BLACKBODY}               &                               &                                 \\
{T$_{in}$ / keV}               &{\bf 0.81\ (0.96$-$0.70)}   &{2.50\ (1.62$-$3.95)}            \\
{intrinsic $N_H$ / 10$^{22}$ cm$^{-2}$} &{\bf 0.00\ ($<$0.11)} &{0.00\ ($<$0.05)}     \\
{log (L$_{X}$ / erg s$^{-1}$)} &{\bf 40.15\ (39.89$-$40.44)}&{40.00\ (39.28$-$41.68)}       \\
{\xs / d.o.f}                  &{\bf 11.3 / 13}                &                                 \\
{CASH statistic / no. of bins} &                               &{  122.5 / 133}                 \\
\hline
\end{tabular}

\caption{Results from simultaneous fitting to the MOS1, MOS2 and pn, 0.5$-$10.0 keV, spectra of 3921-X2 and 3921-X3 (see Figures \ref{3921-blob-spec} and \ref{3921-s3-spec}). The models fitted include a component for Galactic absorption. Abundance is measured relative to solar. The luminosities quoted are unabsorbed, for H$_{0} = 75$ km s$^{-1}$ Mpc$^{-1}$, with limits calculated from the confidence limits on the best-fitting parameters. The errors quoted are the 90\% confidence limits on each single parameter of interest, and the hydrogen column densities refer to the intrinsic absorption, which is fitted in addition to the fixed line-of-sight Galactic absorption. See \S\ref{3921spectxt} for discussion of the fitting.}\label{3921-src-tab} 

\end{center}
\end{table*}

We fit three different model spectra in turn to the 0.5$-$10.0 keV spectra of 3921-X2, the southern source in NGC 3921, extracted from the region shown in Figure \ref{3921-centre-spec}. The MOS1, MOS2 and pn spectra are fit simultaneously, and have a total of 322 counts. The three models are i) an absorbed power-law, frequently used to approximate the X-ray emission from X-ray binaries (XRBs); ii) a MeKaL thermal plasma model, and iii) a disk blackbody model (e.g. Mitsuda et al. 1984; Makishima et al. 1986), which models the X-ray spectrum from a standard accretion disk, consisting of multiple blackbody components. \xmm\ does not have sufficient spatial resolution to determine whether this source is point-like or diffuse, so we fit all these models in an attempt to constrain its nature spectrally. If this extremely bright source {\it is} point-like, then it {\it must} be a ULX,  as it has L$_{X} > 10^{39}$ erg s$^{-1}$ (King et al. 2002).

The results from the spectral fitting are shown in Table \ref{3921-src-tab}. The EPIC spectra, with the best-fitting model spectrum overlaid, are presented in Figure \ref{3921-blob-spec}. The best-fitting model is the disk blackbody model, with the temperature at the inner disk radius, kT$_{inner}$, $=$ 0.81 keV. This is consistent with what one would expect from a black hole binary in the high/soft state (Maccarone et al. 2003). Although fitting the disk blackbody model results in the best fit, it should be noted (from the minimum values of \xs), that neither of the other two models can be rejected. However, the high temperature of the best-fitting MeKaL model argues strongly against a hot gas origin for this source; typical X-ray emitting diffuse gas temperatures in merging systems are $\sim$0.2$-$0.5 keV (Read \& Ponman 1998), and even in virialised systems they only reach $\sim$1.0 keV (Matsushita et al. 1994; Rangarajan et al. 1995).

\subsubsection{3921-X3: the north-western source}

We fit the same three models that we fit to 3921-X2 to the 3921-X3 EPIC spectra. 3921-X3 has fewer counts than 3921-X2 (149 compared with 322), so for this source, the spectra were not grouped, and the Cash statistic was used, rather than \xs\ minimisation. The Cash statistic does not provide an estimation of goodness-of-fit, but we can still compare the relative goodness-of-fit of the different models. 

\begin{figure}
\centerline{\epsfig{file=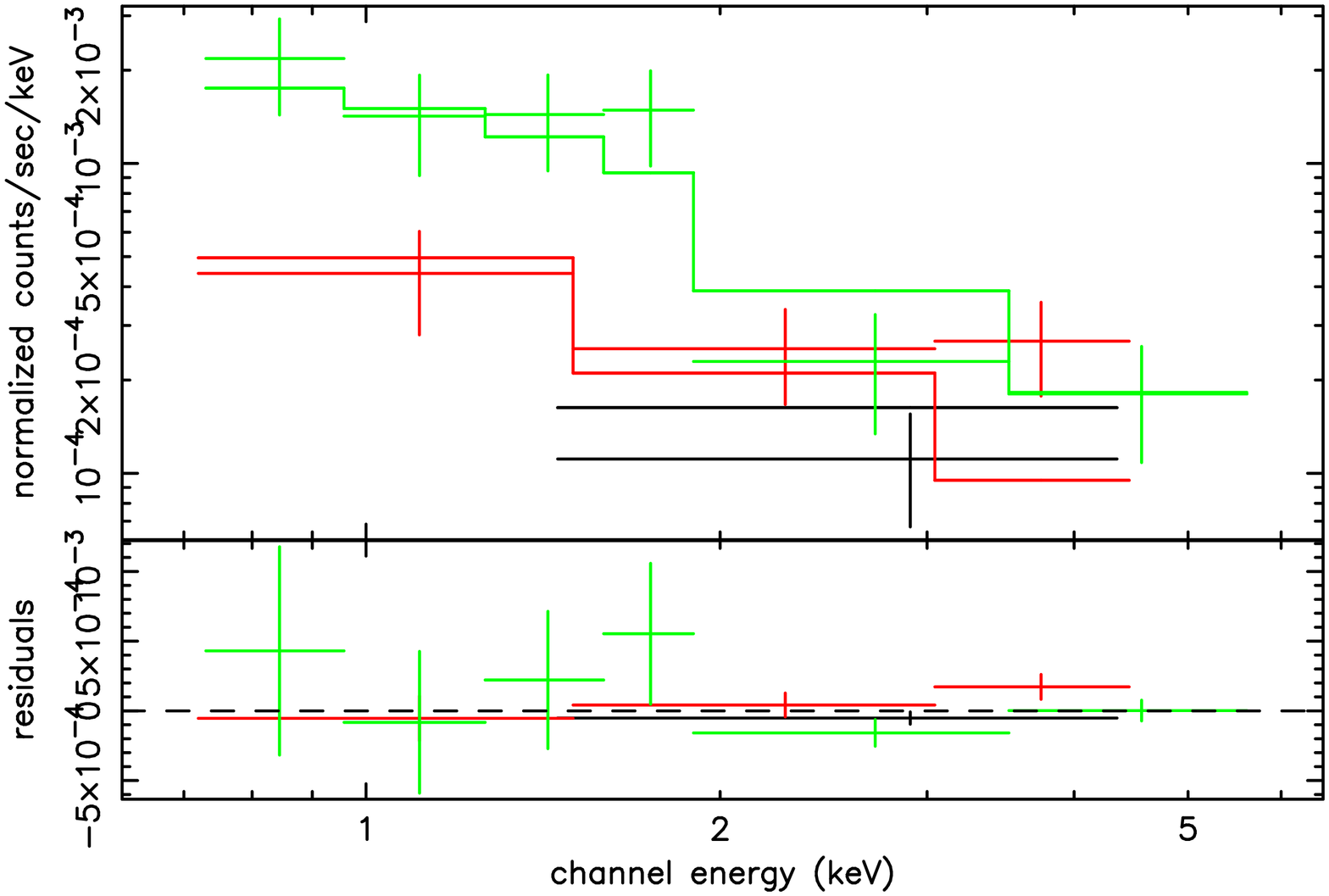,width=6.0cm,angle=-90,clip=}}
\caption{Spectral fit to 3921-X3, the north-western source in NGC 3291. Crosses: pn (upper) and MOS (lower) spectra; solid lines: best-fitting power-law spectra. See Table \ref{3921-src-tab} for details of the models and \S\ref{3921spectxt} for description of the fitting and discussion. The spectra in the plot have been rebinned to at least 10 counts per bin for clarity.}\label{3921-s3-spec}
\end{figure}

The source extraction region, observed spectra and best-fitting model spectra are presented in Figures \ref{3921-centre-spec} and \ref{3921-s3-spec}. The best-fitting parameters for each model are listed in Table \ref{3921-src-tab}. The best-fitting model is the power-law model, with no intrinsic absorption, and an index of 1.28. The other two models cannot be statistically ruled out. However, the extremely high temperature of the best-fitting MeKaL model (63.3 keV) argues very strongly against a hot gas origin for this source. It has a luminosity in excess of 10$^{40}$ erg s$^{-1}$, so, if this is indeed a point source, it must be a ULX.

\section{NGC 7252}

\subsection{Spatial data}\label{7252spattxt}

\begin{figure*}

\centerline{\epsfig{file=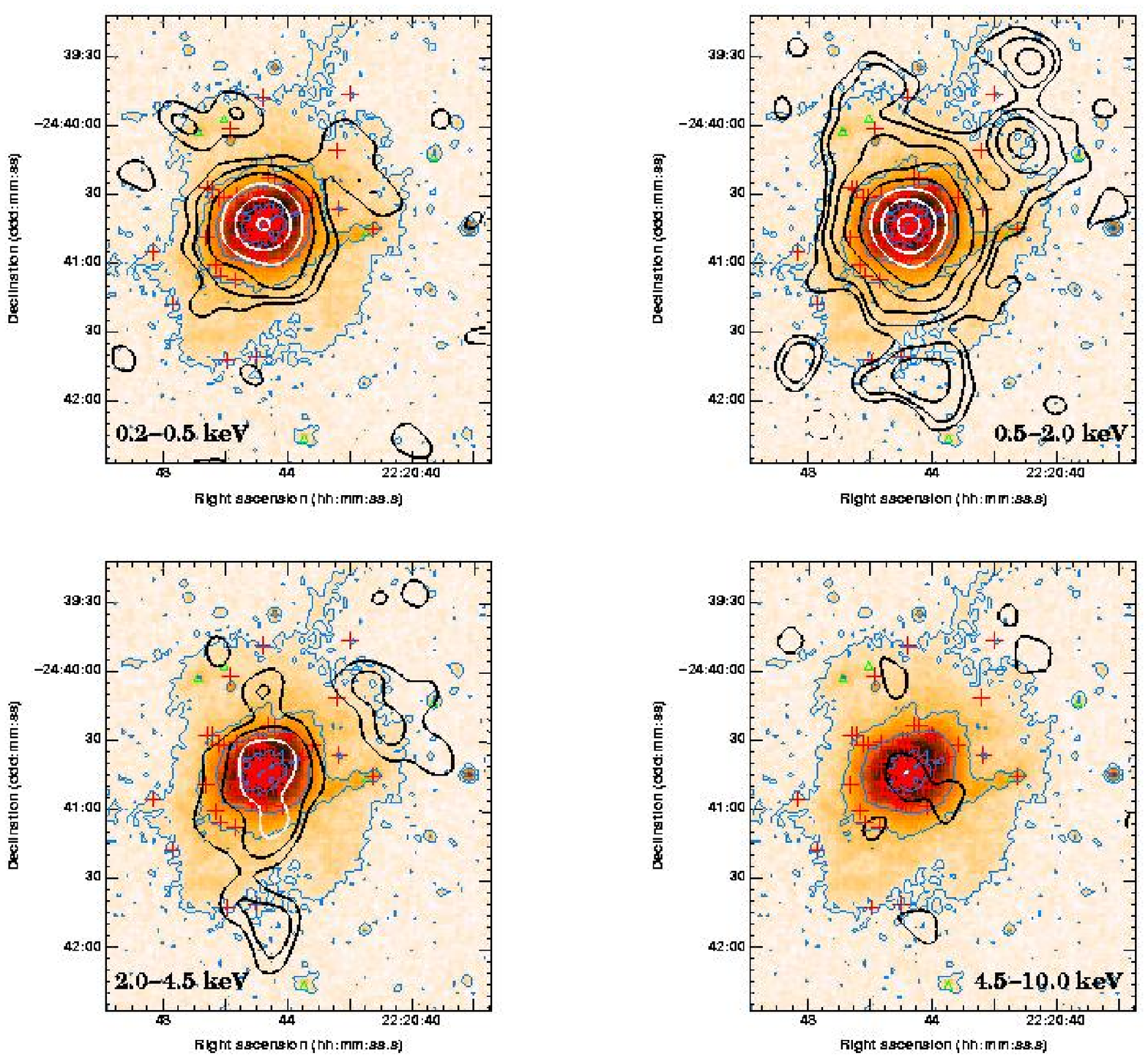,width=16cm,angle=-0,clip=}}

\caption{DSS optical (B$_{j}-$band) images of NGC 7252 (greyscale with thin grey contours) with all-EPIC X-ray emission contours (thick black and white, as described in $\S$ 2) overlaid. Top left: 0.2$-$0.5 keV; top right: 0.5$-$2.0 keV; bottom left: 2.0$-$4.5 keV; bottom right: 4.5$-$10.0 keV. The X-ray contour levels increase by factors of two from 1.125x10$^{-6}$ counts s$^{-1}$. Crosses: globular clusters and star clusters (Miller et al. 1997); triangles: galaxies (Maddox et al. 1990). The peak of the nuclear X-ray emission is aligned with the centre of the optical emission. Three off-nuclear sources are clearly detected, to the north-west, south-south-west and south. There is very little X-ray emission detected above $\sim$ 4.5 keV. The dashed circle in the top right plot represents the half-energy width (14\arcsec). The pixel sizes are 2.5\arcsec\ by 2.5\arcsec.}\label{7252ebands}
\end{figure*}

In Figure \ref{7252ebands}, the background-subtracted, exposure-corrected, smoothed X-ray contours for NGC 7252 in four different energy bands (0.2$-$0.5, 0.5$-$2.0, 2.0$-$4.5 and 4.5$-$ 10.0 keV) are presented, overlaid on the optical (B$_{j}-$band) DSS image. Globular and star clusters are marked with crosses. The X-ray emission is fairly symmetrical, with no shift in emission with energy. The optical and X-ray nuclei are well-aligned. The globular / star clusters are centrally concentrated. There are more of them and they are more luminous than those in NGC 3921 (Miller et al. 1997 and Table \ref{props}). There is negligible emission above $\sim$ 4.5 keV. The central hot diffuse emission is not very extended in any of the energy bands.

The CIAO tool {\em wavdetect} was used to search for point-like
sources in the 28 ksec \chandra\ archival observation. Scales from
1$-$16 pixels (0.5$-$8\arcsec) were investigated. In addition to the
extended nuclear emission, a total of 6 bright, high-significance ($>
4\sigma$) point (i.e. small-scale) sources were detected in the
0.3$-$7 keV band within or close to the optical confines of NGC
7252. Seven lower-significance (between $2.3\sigma$ and $4\sigma$)
point sources were also detected. A further source (X2) is obvious
close to the bright nucleus, though not formally detected by the
software (probably because of its proximity to the
nucleus). Properties of this source have been estimated by calculating
by hand the counts within an ellipse around this source, with the size
and shape of the ellipse estimated by comparison with the surrounding
sources: 7252-X1, X3, X9 and X10. The {\it wavdetect} intensities were
checked by performing aperture photometry using custom-defined
circular apertures of 1$-$8 pixel radius (1 pixel $\sim$
0.5\arcsec). The {\it wavdetect}ed counts were, within the
uncertainties, equal to those obtained within small ($\sim$ 1$-$2
pixels) radii, except for the nucleus, which is extended and therefore required a larger ($\sim$ 4.5 pixels) radius.

\begin{figure}
\begin{center}
\centerline{\epsfig{file=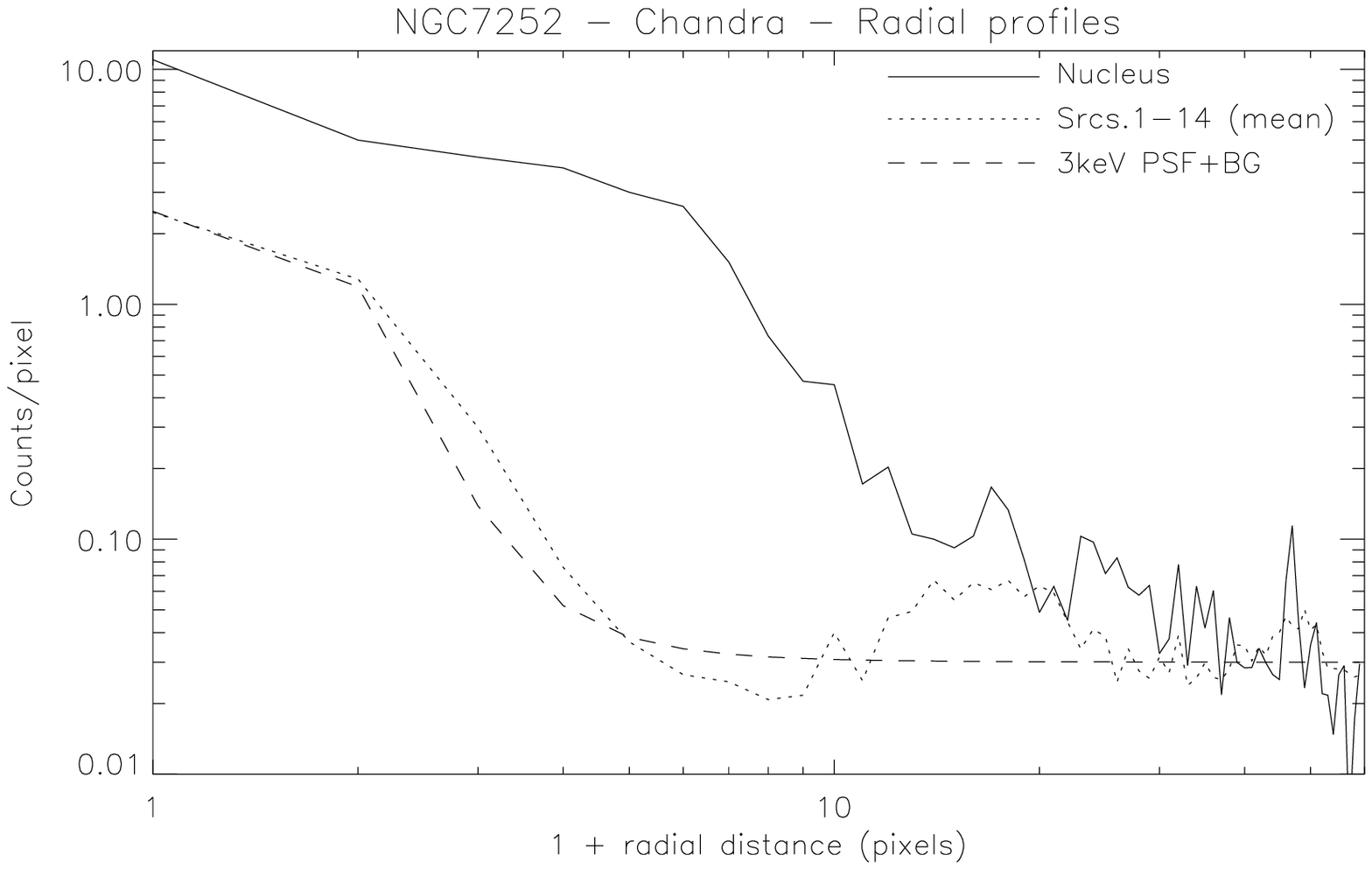,width=9.5cm,angle=-0,clip=}}
\caption{Radial profiles of i) the nucleus (solid line), ii) the mean
of the sources 7252-X1$-$14 (except 7252-X8, which is not detected by
\chandra) dotted line, and iii) a 3 keV on-axis ACIS PSF, scaled to
the mean off-nuclear source profile, to which the estimated background
rate is added (dashed line). The mean off-nuclear source profile is
point-like, whereas the nucleus is extended beyond the
PSF.}\label{chradprof}
\end{center}
\end{figure}

The point-like nature of the sources is confirmed by comparing the
\chandra\ radial profiles with the PSF, as shown in Figure
\ref{chradprof}. Here, the radial profiles of i) the nucleus, ii) the
mean of the sources 1$-$14 (except 7252-X8, which is not detected by
\chandra), and iii) a 3 keV on-axis ACIS PSF, created using
CIAO-mkpsf and scaled to the mean off-nuclear source profile, to which
the estimated background rate is added. We take the mean profile of
the off-nuclear sources, as the individual count rates are too low to
justify comparison with the PSF. The mean non-nuclear source profile
follows the PSF distribution and is therefore point-like. The nucleus,
however, is definitely extended, out to 5\arcsec and beyond. 

\begin{figure}
\begin{center}
\centerline{\epsfig{file=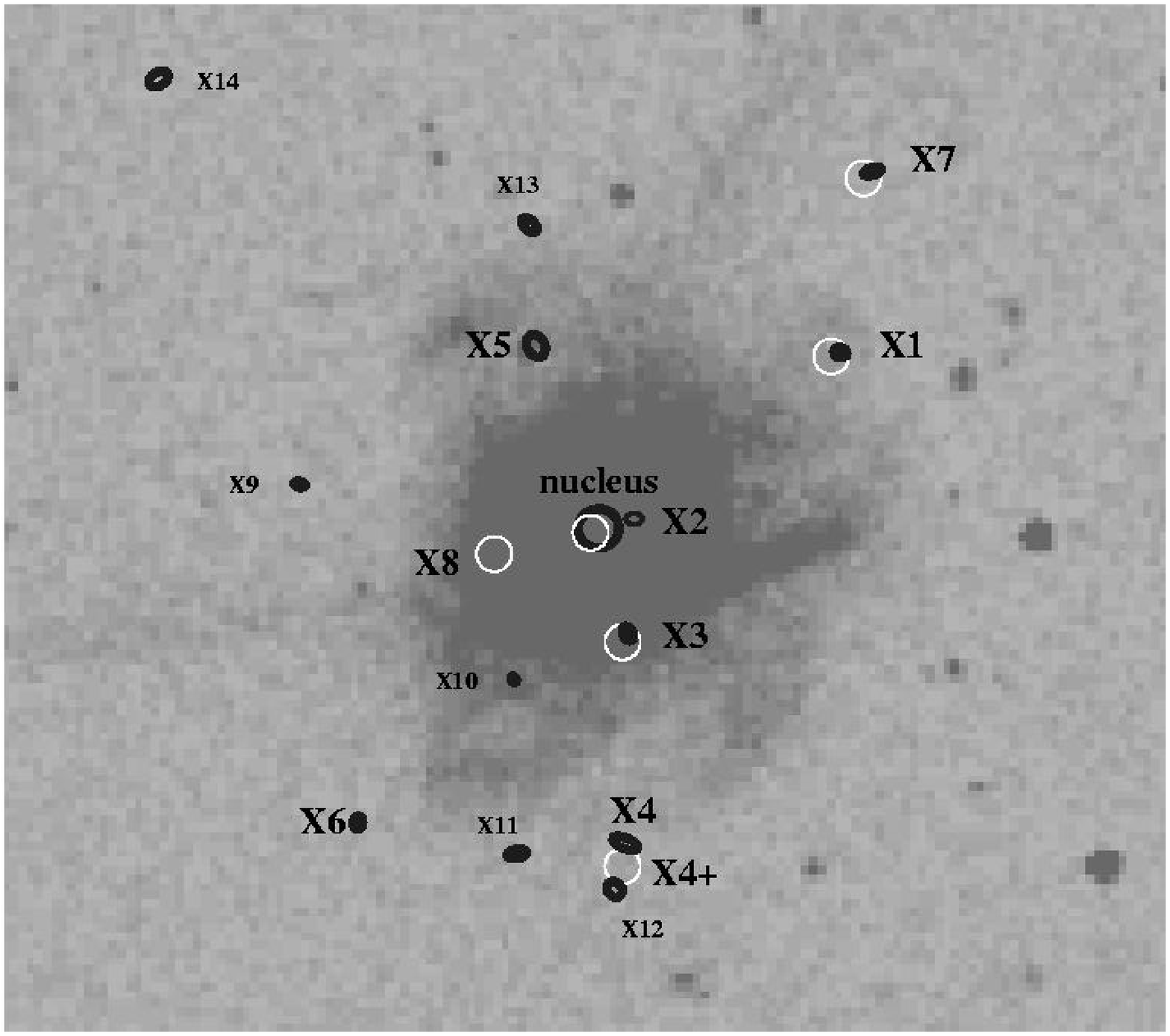,width=7.5cm,angle=-90,clip=}}
\caption{Sources detected in the \xmm\ (white circles) and \chandra\ (black ellipses) observations of NGC 7252, overlaid on the DSS (B$_{j}$) image. Large fonts refer to those sources detected with $>$4$\sigma$ significance, and small fonts to those with 2.3$< \sigma <$4 significance. 7252-X5 is a foreground star.}\label{chandra} 
\end{center}
\end{figure}

The sources are listed in Table \ref{chandrasrcs} and plotted in
Figure \ref{chandra}. 7252-X2 is the only source which may be
associated with an observed globular or star cluster. 7252-X5 is a
foreground star, and we exclude it from further discussion.

In the \xmm\ data, using the SAS source-searching software, we identify, in addition to the nucleus, five off-nuclear X-ray sources within the body of the galaxy from these data (see Table \ref{7252srcs}). The whole area covered by the \chandra\ observation was searched. The minimum detection likelihood for inclusion of a source in the output list is 10, so that the probability that a detected source is random Poissonian noise is 4.5x10$^{-5}$. 

None of these sources is obviously associated with any observed globular or star clusters. Although the spatial resolution of \xmm\ is not sufficient to determine whether these sources are point-like or extended, two of the five \xmm\ sources (7252-X1 and X3) are confirmed (with $\sigma >$ 4.0) as point sources by the \chandra\ detections. 7252-X7 is also identified as point source by the \chandra\ observation, but with lower significance ($\sigma =$ 2.7). The co-ordinates of these sources for which there are detections in both the \xmm\ and \chandra\ data agree very well. However, \xmm\ source 7252-X4$+$ lies between 7252-X4 and 7252-X12. It most likely includes contributions from the (low-significance) sources 7252-X12 and 7252-X11. Hence, we label it 7252-X4$+$ to differentiate it from the \chandra\ source 7252-X4.

This brings us to a total of seven significant ultra-luminous off-nuclear X-ray sources in NGC 7252 (7252-X1$-$X8, excluding 7252-X5), six of which we know to be point-like. There are in addition a further six point sources detected only in the \chandra\ data with lower (2.3$< \sigma <$4) significance. 

\begin{table*}

\begin{center}
\begin{tabular}{llllll}

\hline
source  & RA(J2000) / h:m:s & Dec(J2000) / d:m:s & significance & counts & log (L$_{X}$ / erg s$^{-1}$) \\
\hline
	   
{\bf NUCLEUS}  & {\bf 22:20:44.76}  & {\bf -24:40:41}  & {\bf 25.0 }    &256.89$\pm$18.5  & {\bf 40.75\ (40.71$-$40.78)} \\
{\bf 7252-X1}  & {\bf 22:20:41.15}  & {\bf -24:40:07}  & {\bf 8.8  }    &18.54$\pm$4.36   & {\bf 39.60\ (39.49$-$39.70)} \\
{\bf 7252-X2}  & {\bf 22:20:44.23}  & {\bf -24:40:43}  & {\bf $\sim$6.7}&15.0 $\pm$4.0    & {\bf 39.52\ (39.38$-$39.62)} \\
{\bf 7252-X3}  & {\bf 22:20:44.32}  & {\bf -24:41:04}  & {\bf 15.5  }   &34.27$\pm$5.92   & {\bf 39.88\ (39.79$-$39.94)} \\
{\bf 7252-X4}  & {\bf 22:20:44.36}  & {\bf -24:41:47}  & {\bf 6.7   }   &15.15$\pm$4      & {\bf 39.52\ (39.51$-$39.53)} \\
{\bf 7252-X5}  & {\bf 22:20:45.70}  & {\bf -24:40:05}  & {\bf 8.9   }   &21.67$\pm$4.80   & --- \\
{\bf 7252-X6}  & {\bf 22:20:48.36}  & {\bf -24:41:42}  & {\bf 5.0   }   &10.53$\pm$3.32   & {\bf 39.36\ (39.20$-$39.48)} \\
7252-X7  & 22:20:40.67  & -24:39:30  & 2.7    &5.58$\pm$2.45   & 39.08\ (38.83$-$39.24) \\
7252-X8  &  ---         &  ---       & ---    &     & ---   \\
7252-X9  & 22:20:49.23  & -24:40:33  & 2.3    &4.73$\pm$2.23   & 39.00\ (38.73$-$39.18) \\
7252-X10 & 22:20:46.03  & -24:41:13  & 2.9    &5.74$\pm$2.45   & 39.11\ (38.86$-$39.25) \\
7252-X11 & 22:20:45.98  & -24:41:49  & 2.4    &4.74$\pm$2.24   & 39.00\ (38.73$-$39.18) \\
7252-X12 & 22:20:44.52  & -24:41:56  & 3.3    &7.33$\pm$2.83   & 39.20\ (38.99$-$39.34) \\
7252-X13 & 22:20:45.79  & -24:39:41  & 3.2    &6.59$\pm$2.65   & 39.15\ (38.93$-$39.30) \\
7252-X14 & 22:20:51.34  & -24:39:11  & 2.8    &5.72$\pm$2.45   & 39.11\ (38.85$-$39.25) \\
\hline

\end{tabular}

\caption{Sources detected in the \chandra\ NGC 7252 data set, using {\em wavdetect}. The positions of the sources are plotted in Figure \ref{chandra}. Sources detected with greater than 4$\sigma$ significance are printed in bold type. The luminosities are calculated in the 0.5$-$ 10.0 keV range, using a power-law model with an index of 1.5, and Galactic line-of-sight absorption. The unabsorbed luminosities are quoted. H$_{0} =$ 75 km s$^{-1}$Mpc$^{-1}$. 7252-X5 is a foreground star.}\label{chandrasrcs} 

\end{center}
\end{table*}

\begin{table*}

\begin{center}
\begin{tabular}{lllll}

\hline
source  & RA(J2000) / h:m:s & Dec(J2000) / d:m:s & detection likelihood & 10$^{-2}$ counts s$^{-1}$ \\
\hline
NUCLEUS    &     22:20:44.79 & -24:40:43 &  5109.8 &   6.392$\pm$0.173 \\
7252-X1    &     22:20:41.20 & -24:40:07 &  108.7  &   0.499$\pm$0.071 \\
7252-X2    &     ---         & ---       & ---     &    ---    \\
7252-X3    &     22:20:44.45 & -24:41:06 &  144.1  &   0.858$\pm$0.091 \\
7252-X4$+$&     22:20:44.48 & -24:41:51 &  83.9   &    0.461$\pm$0.068 \\
7252-X5    &     ---         &  ---      & ---     &    ---    \\
7252-X6    &     ---         &  ---      & ---     &    ---    \\
7252-X7    &     22:20:40.79 & -24:39:31 &  26.2   &   0.227$\pm$0.060 \\
7252-X8    &     22:20:46.41 & -24:40:48 &  15.9   &   0.300$\pm$0.073 \\
\hline

\end{tabular}

\caption{Sources detected in the \xmm\ NGC 7252 data set, using the SAS source-searching software. The positions of the sources are plotted in Figure \ref{7252hr}. See also Figure \ref{chandra} for the \chandra-identified sources. 7252-X5 is a foreground star.}\label{7252srcs} 

\end{center}
\end{table*}

\begin{figure}
\centerline{\epsfig{file=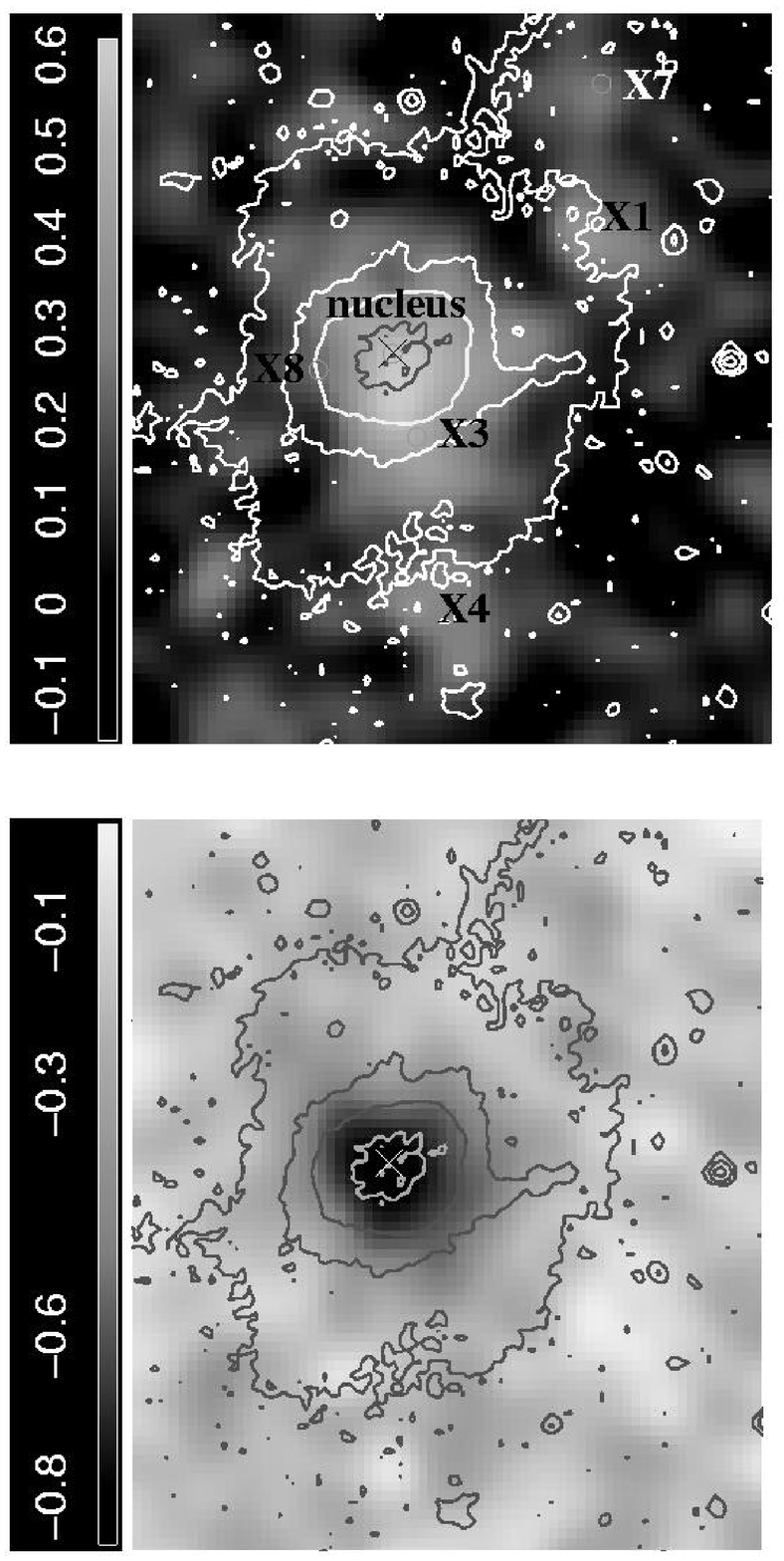,width=7.5cm,angle=-0,clip=}}
\caption{Hardness ratio maps for the all-EPIC observation of NGC 7252, with optical contours from the DSS image (B$_{j}-$band) overlaid. Top: HR$_{1}$ (soft); bottom: HR$_{2}$ (medium). See $\S$ 2 for details of the hardness ratio calculation.  Lighter grey represents harder emission. The cross locates the centre of NGC 7252 (RA(J2000) 22h20m44.775s, Dec(J2000) -24d40m41.83s; Miller et al. 1997), and the circles the X-ray sources detected in the galaxy, which are labelled in the top plot as in Table}\label{7252hr}
\end{figure}

Figure \ref{7252hr} shows the two hardness ratio (HR) maps calculated
for NGC 3291 as described in $\S$ \ref{obsanddr}, with the contours
from the DSS optical (B$_{j}-$band) image superposed. In these plots,
darker shades represent softer emission, and lighter shades are
harder. It should be noted that the grey scales in each plot are not
the same. There is no HR$_{3}$ map because there are too few photons
detected in the hardest energy band. Most emission is in 0.5$-$2.0 keV
range; the HR$_{1}$ map shows the hardness peaking in the centre of the
galaxy. In HR$_{2}$ the hardness ratio declines towards the centre. The
positions of the five (\xmm) off-nuclear sources are over-plotted on the HR$_{1}$
map.

\begin{figure}
\centerline{\epsfig{file=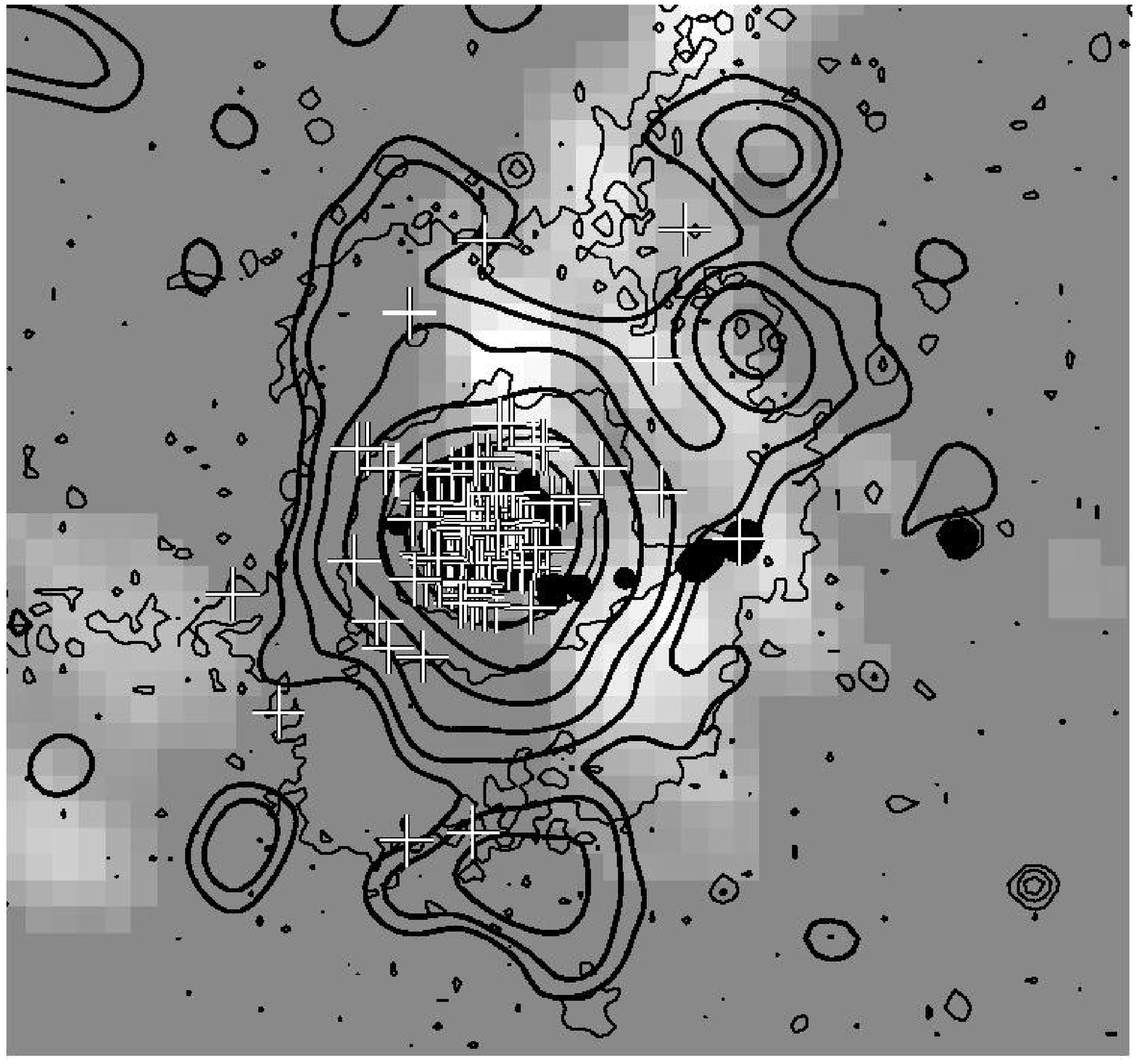,width=9.0cm,angle=-0,clip=}}
\caption{Hot, warm and cold gas in NGC 7252. Thick contours: all-EPIC, 0.5$-$2.0 keV emission, as described in $\S$ 2. black patches: H$\alpha$ (Hibbard et al. 1994); thin contours: DSS optical (B$_{j}-$band) image; white-on-grey scale: HI (Hibbard et al. 1994); crosses - globular clusters and star clusters (Miller et al. 1997). }\label{7252Hregs}
\end{figure}

Figure \ref{7252Hregs} shows the H$\alpha +$[NII] (KPNO 2.1m
telescope) and HI (VLA) emission detected in NGC 7252 (Hibbard et
al. 1994). H$\alpha +$[NII] is mostly concentrated in the nucleus,
with a loop extending out to the west. HI is associated only with the
tails, both to the north-west and to the east. None of the off-nuclear
X-ray sources is associated with the H$\alpha$ region, except 7252-X2
(Figure \ref{chandra}), which lies within the central H$\alpha$
region. The lop-sidedness of the HI emission is not reflected in the
X-ray emission. 7252-X1, X2 and X7 are spatially correlated with the
region where the north-western HI arm comes across the front of the
remnant body (Hibbard and van Gorkom 1996), but we cannot tell
whether they lie within, in front of, or behind the HI region.

\begin{figure}
\centerline{\epsfig{file=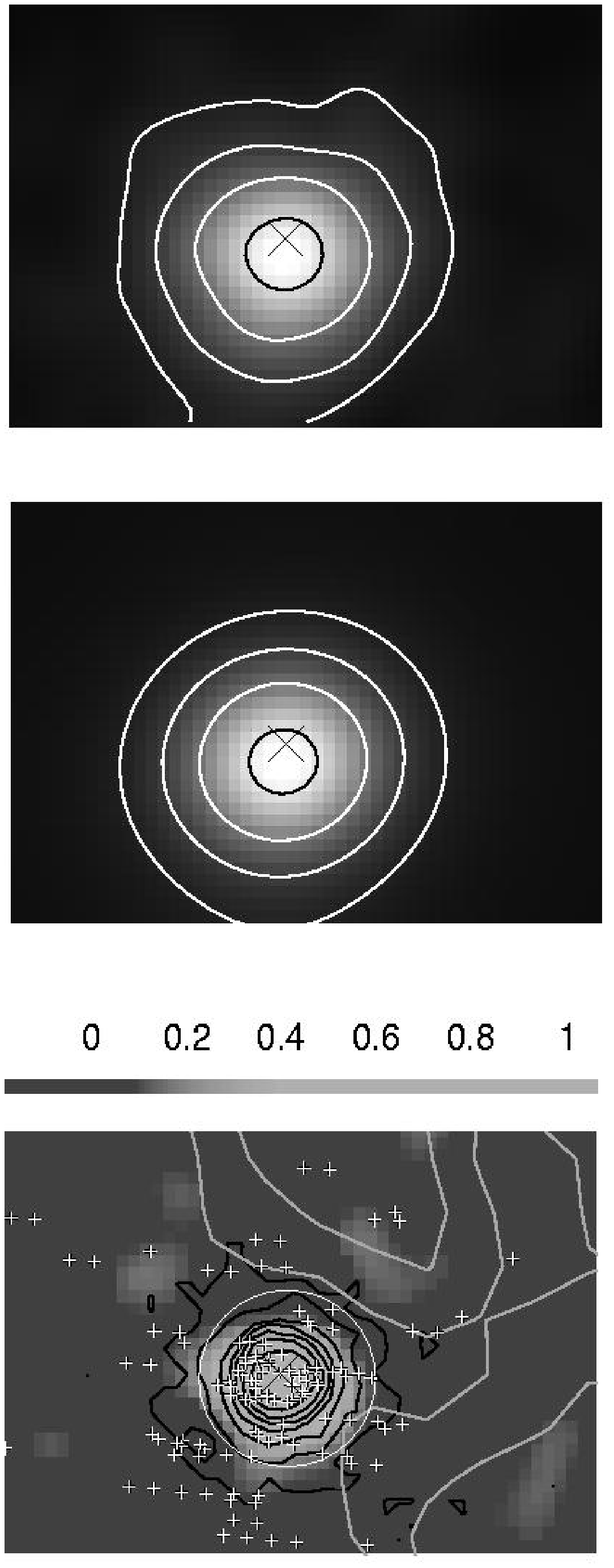,width=6.5cm,angle=-0,clip=}}
\caption{Spatial fitting to the 0.3$-$1.2 keV, MOS1, MOS2 and pn mosaiced image (top) of the main body of NGC 7252. Top: mosaiced image, with contour levels overlaid; middle: best-fitting 2-D $\beta$-model, plus constant background, with the same contour levels overlaid; bottom: residuals ($data-model$) and colour bar, where lighter colours are positive, and darker colours are negative. The units are counts. The image and residual plots have been simply-smoothed with a Gaussian of FWHM 2.75\arcsec, after the fitting has been carried out. Overlaid are globular clusters and stellar associations (crosses; Miller et al. 1997), H$\alpha$ contours (dark lines) and HI contours (light grey lines). The diagonal cross marks the optical centre of NGC 7525. See \S\ref{7252spattxt} for details of the fitting. The circle marks the nuclear region excluded from the fit. Each plot covers the same region, of size 58.3\arcsec x40.7\arcsec, centred on the nucleus.}\label{7252spat}
\end{figure}

We model the surface brightness distribution of the main body of NGC 7252, within a rectangle 58.3\arcsec\ by 40.7\arcsec, centred on the nucleus, following the same prescription as for NGC 3291. The energy range is restricted to 0.3$-$1.2 keV, in order to exclude contributions from the power-law component (see \S\ref{7252spectxt} and the spectrum in Figure \ref{7252-centre-spec}). The central region hosts at least one point source (7252-X2), so this is masked out of the fit (see Figure \ref{7252spat}). Table \ref{7252spatfittab} lists the best-fitting parameters for the $\beta$-model, and Figure \ref{7252spat} plots the data, best-fitting model and fit residuals ($data-model$). Again, the HI and H$\alpha$ emission contours are over-plotted on the residuals image. Unlike the marked correspondence in the main body of NGC 3921, in NGC 7252, there is no correlation between the spatial distribution of HI and the deficit or excess of X-ray emission. However, the X-ray excess, masked out of the fit, is in the very central region where the H$\alpha$ peaks. Candidate globular clusters and stellar associations (Miller et al. 1997) are marked on the residuals plot, but these too do not obviously correspond with excess (or deficit) emission (except in the case of the masked-out central region). The power-law index, $\alpha =$ 1.31$\pm$0.02, is less steep than that for NGC 3921, and NGC 7252 has a much more compact core, with a core radius 1.05$\pm$0.03 kpc. As with NGC 3921, this model should be interpreted as a description of the shape of the hot diffuse gas surface brightness distribution, rather than as a physical model of it. Source 7252-X8 is most likely responsible for the excess of emission to the east and slightly south of the nucleus.

\begin{figure}
\centerline{\epsfig{file=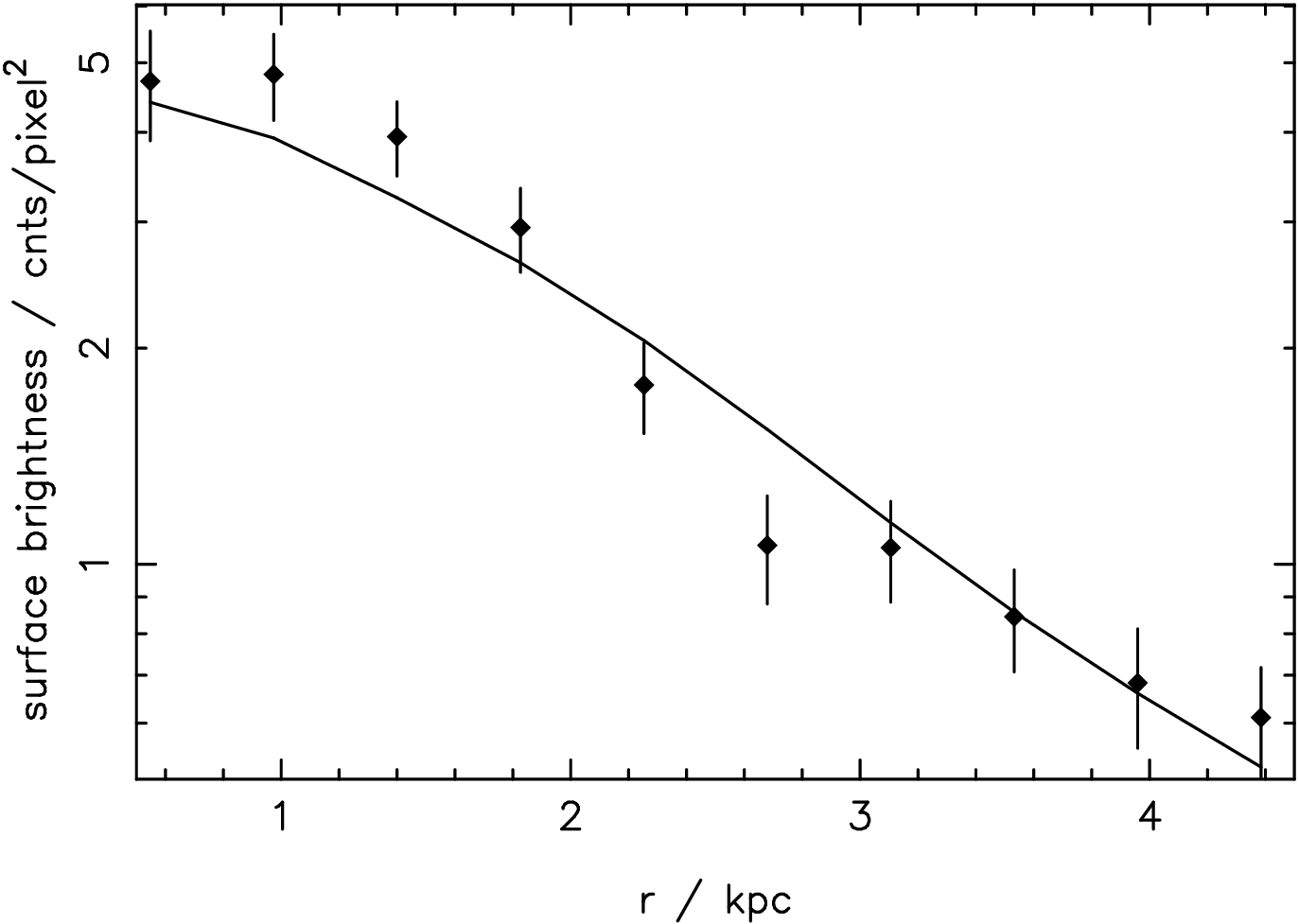,width=6.5cm,angle=-90,clip=}}
\centerline{\epsfig{file=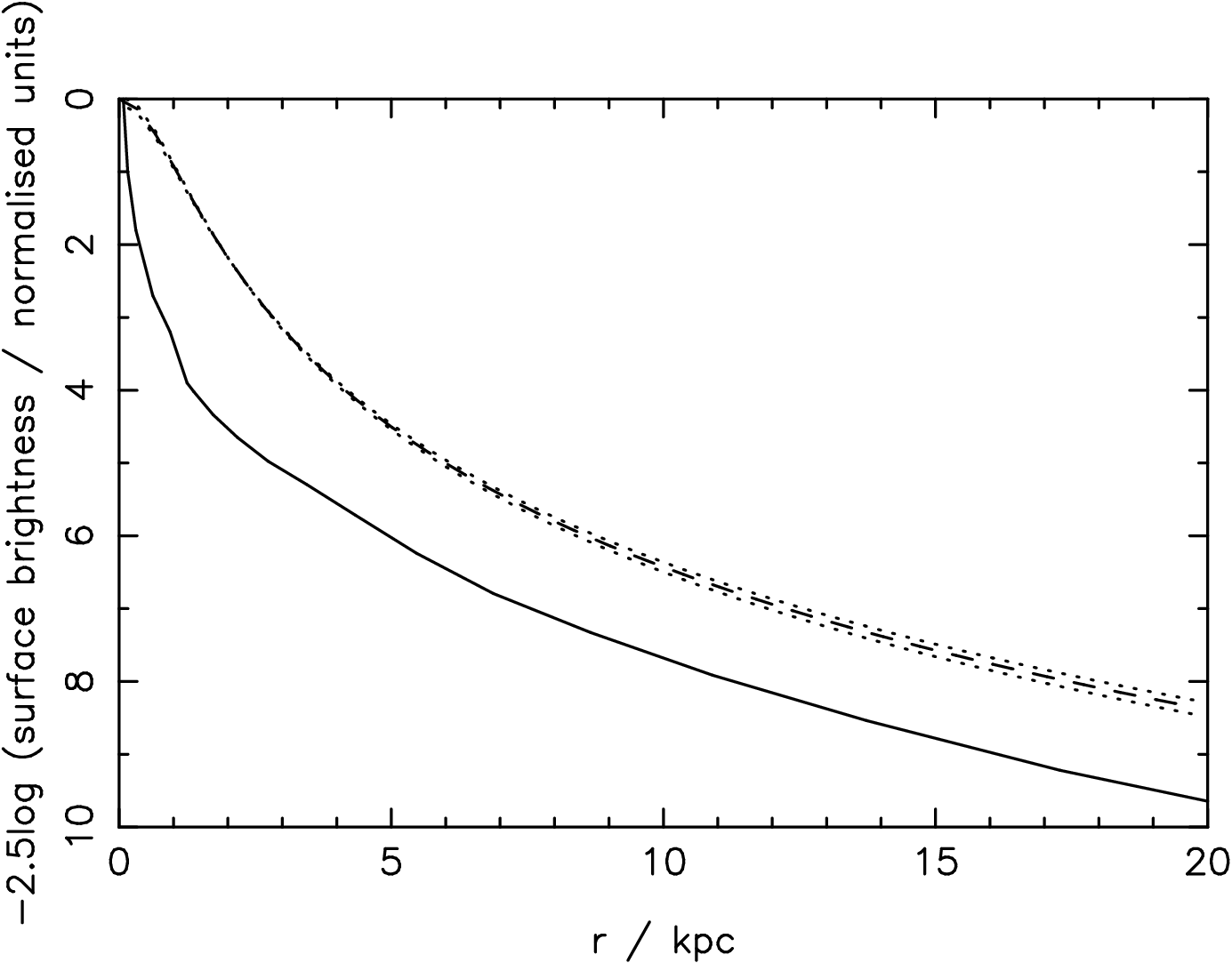,width=7.0cm,angle=-90,clip=}}
\caption{Top: diamonds: radial profile (raw data, binned over circular annuli) of the diffuse gas in NGC 7252, plus error bars, over the range of the 2-dimensional fitting; solid line: model radial profile, using the best-fitting 2-dimensional model parameters, convolved with the PSF. Bottom: dashed line: model radial profile of the hot diffuse gas in NGC 7252, using the fitted parameters from the best-fitting 2-D spatial fitting (see Table \ref{7252spatfittab} and \S\ref{7252spattxt}). The dotted lines represent the uncertainties in the model fitting. Solid line: mean V-band surface brightness profile (Schweizer 1982, Miller et al. 1997). The surface brightness has been normalised to unity at r$=$0 in both cases.}\label{7252sbr}
\end{figure}

Figure \ref{7252sbr} compares the modelled X-ray surface brightness distribution with the mean V-band radial profile. The hot diffuse gas emission has a compact central core, although, as for NGC 3921, it is not as compact as the central V-band radial profile. The X-ray surface brightness distribution has a similar slope to that of the optical surface brightness distribution at larger radii (\gs4 kpc). The raw X-ray data, together with the best-fitting model surface brightness distribution, are also shown in Figure \ref{7252sbr}.

\begin{table}

\begin{center}
\begin{tabular}{cccc}

\hline
object   & core radius / kpc & $\alpha$  & amplitude / cnts \\\hline
NGC 7252 & 1.05$\pm$0.03 & 1.31$\pm$0.02 & 15.2$\pm$0.8  \\
\hline
\end{tabular}

\caption{The results of 2-D spatial fitting to the 0.3$-$10.0 keV, MOS1, MOS2 and pn mosaiced images of the main body of NGC 7252. The fitted model is a $\beta$-model, as described in \S\ref{3921spattxt}. $r_{0}$ is calculated for H$_{0} =$ 75 km s$^{-1}$Mpc$^{-1}$.  Figure \ref{7252spat} presents the data, best-fitting models and residuals of the fits. See \S\ref{7252spattxt} for discussion.}\label{7252spatfittab}
\end{center}
\end{table}

\subsection{Spectral fitting}\label{7252spectxt}

\subsubsection{Nucleus}

\begin{table*}
\begin{center}

\begin{tabular}{lllll}

\hline
 {{  model}}         & {  MeKaL 1   }             & {  MeKaL 2   }            & {  power-law } & {  total     }  \\
\hline
 {kT / keV}        & {  0.72\ (0.66$-$0.77)} & {  0.36\ (0.33$-$0.41)}& {  }           & {  }            \\
 {abundance }   & {  0.14\ (0.12$-$0.16)} & {  0.64\ (0.54$-$0.72)}& {  }           & {  }            \\
 {photon index}    & {  }                       & {  }                      & {  1.72\ (1.66$-$2.39)} & {  }\\  
 {intrinsic $N_H$ / 10$^{22}$ cm$^{-2}$}    & {  }       & {  }                      & {  1.71\ (1.26$-$3.66)} & {  }\\ 
 {log (L$_{X}$ / erg s$^{-1}$)}    & {  40.12\ (40.07$-$40.15)}  & {  39.89\ (39.80$-$39.98)}                 & {  40.28\ (40.06$-$40.31) }     & {  40.60\ (40.47$-$40.64) }      \\
 {\xs  / d.o.f}    & {  }                       & {  }                      & {  }           & {  58.4 / 46}   \\
\hline

\end{tabular}

\caption{Results from simultaneous fitting to the MOS1, MOS2 and pn, 0.5$-$10.0 keV, spectra of the nucleus of NGC 7252 (see Figure \ref{7252-centre-spec}). The model fitted has components for the Galactic extinction, two MeKaL components, and an absorbed power-law. Abundance is measured relative to solar. The luminosities quoted are unabsorbed, for H$_{0} = 75$ km s$^{-1}$ Mpc$^{-1}$. The errors quoted are the 90\% confidence limits on each single parameter of interest, and the hydrogen column density refers to the intrinsic absorption, which is fitted in addition to the fixed line-of-sight Galactic absorption. See \S\ref{7252spectxt} for description of the fitting and discussion. }\label{7252-centre-tab} 
\end{center}
\end{table*}

\begin{figure}
\centerline{\epsfig{file=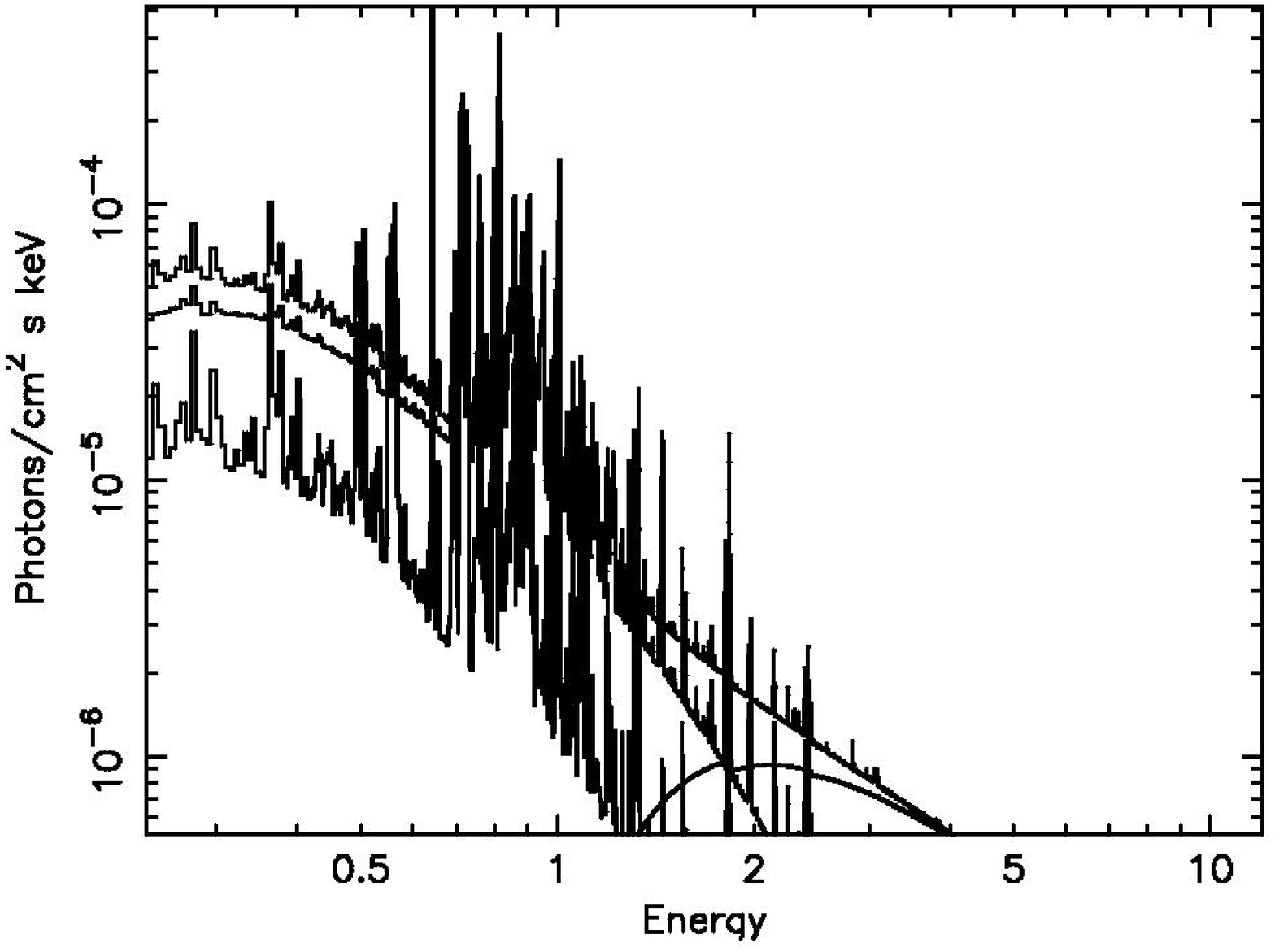,width=6.5cm,angle=-90,clip=}}
\centerline{\epsfig{file=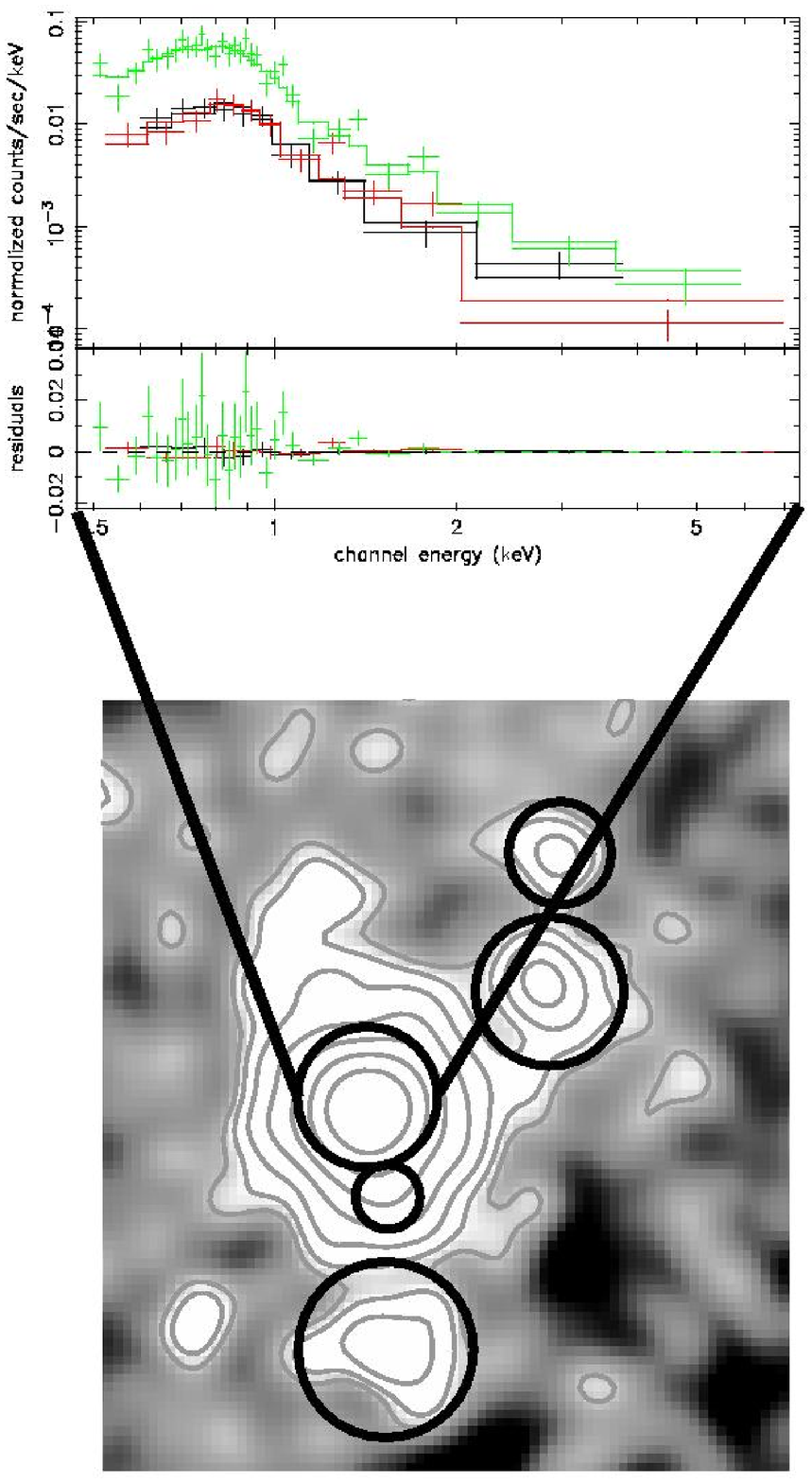,width=10cm,angle=-0,clip=}}
\caption{Spectral fit to the nucleus of NGC 7252. Top: Spectral plot of the best-fitting model (top line) showing the two MeKaL components, which dominate the model, and the modest absorbed power-law component, which peaks at $\sim$ 2 keV. Middle: crosses: pn (upper) and MOS (lower) spectra; solid lines: best-fitting model spectra. See Table \ref{7252-centre-tab} for details of the models. Bottom: X-ray contour map (0.5$-$2.0 keV) showing the nuclear region from which the spectra were extracted. See \S\ref{7252spectxt} for description of the fitting and discussion. }\label{7252-centre-spec}
\end{figure}

The 0.5$-$10.0 keV MOS1, MOS2 and pn spectra of NGC 7252, totalling 1375 counts, extracted from the region shown in Figure \ref{7252-centre-spec}, were modelled simultaneously by a two-component MeKaL model plus an absorbed power-law component. The best-fitting parameters of this model are listed in Table \ref{7252-centre-tab}. Figure \ref{7252-centre-spec} shows a spectral plot of the various contributions from the model components. The power-law component is much less luminous than in NGC 3921, and there is negligible emission above 4.0 keV. The goodness-of-fit of the best-fitting model is acceptable. The hot diffuse gas components in NGC 7252 are slightly cooler than those in NGC 3921. The luminosity of the nucleus estimated from the \chandra\ data set does not agree well with the \xmm\ results, but this is a result of the different models fitted; fitting a power-law with a slope of 1.5 to the \xmm\ data results in the same luminosity as for the \chandra\ data, within the uncertainties. We are able to fit a much more complex model to the \xmm\ spectra, as there are far more counts available than in the \chandra\ data. This gives us an acceptable goodness-of-fit, which is not the case with the simple power-law model.

Awaki et al. (2002) speculate that the hard (2.0$-$10 keV) X-ray emission that they detect in their 43.5 ks observation of this galaxy using the Solid-State Imaging Spectrometer on ASCA (1998) indicates the existence of nuclear activity or an intermediate-mass black hole in the remnant body. However, there is no other evidence of nuclear activity in this well-studied object. The ASCA spectrum is extracted from a circular region, 6\arcmin\ in diameter, centred on the nucleus. This region includes all of the sources we detect, so it seems more likely that the hard X-ray component in the ASCA spectrum represents the emission from unresolved X-ray point sources. The total X-ray luminosity observed by Awaki et al., 8.1x10$^{40}$ erg s$^{-1}$, is consistent, within the uncertainties, with the sum of the luminosities of all the sources we detect.

\subsubsection{Off-nuclear sources}\label{7252onsrcs}

\begin{table*}
\begin{center}
\begin{tabular}{lllll}

\hline
{SOURCE}                      &  {\bf 7252-X1}                & {\bf 7252-X3}                  & {\bf 7252-X4$+$}                 &  {\bf 7252-X7}        \\
{counts     }                 &{221}                          & { 109 }                        & {  264 }                         & {79}            \\
\hline
{POWER-LAW}                   &                               &                                &                                  &                 \\ 
{photon index}                &{\bf 1.38\ (1.10$-$1.63) }  & {\bf 1.98\ (1.58$-$2.29) }  & {1.88\ (1.42$-$2.14)  }       &{\bf 2.06\ (1.20$-$3.32)  }  \\ 
{intrinsic $N_H$ / 10$^{22}$ cm$^{-2}$}&{\bf 0.00\ ($<$0.08)}   & {\bf 0.00\ ($<$0.06)}   & {0.20\ (0.05$-$0.33)}   &{\bf 0.00\ ($<$0.19)}   \\ 
{log (L$_{X}$ / erg s$^{-1}$)}&{\bf 39.89\ (39.76$-$40.05)}& {\bf 39.54\ (39.45$-$39.71)}& {39.89\ (39.80$-$40.10)}      & {\bf 39.14\ (38.54$-$39.95)} \\
{CASH statistic / no. of bins}& {\bf 184.4 / 204}             &     {\bf 109.4 / 101}          & {241.4 / 232}                    & {\bf 64.1 / 78} \\
\hline
{DISK BLACKBODY}               &                              &                                &                                  \\
{T$_{in}$ / keV}               &{1.70\ (1.21$-$2.67) }     &   {1.24\ (0.70$-$1.59)}     & {\bf 1.30\ (1.04$-$2.32) }    & {0.42\ (0.24$-$2.90) }   \\
{intrinsic $N_H$ / 10$^{22}$ cm$^{-2}$} & {0.00\ ($<$0.07)}     &{0.00\ ($<$0.02)}  & {\bf 0.00\ ($<$0.11) }    & {0.00\ ($<$0.22) }    \\
{log (L$_{X}$ / erg s$^{-1}$)} &{39.80\ (39.18$-$40.54)}   &{39.56\ (38.52$-$39.99)}     & {\bf 39.73\ (39.33$-$40.71)}  & {38.88\ (37.46$-$42.04)}  \\
{CASH statistic / no. of bins} &  {187.0 / 204}                &   {124.4 / 101}               & {\bf 240.3 / 232}                & {66.6 / 78}                 \\
\hline
\end{tabular}

\caption{Results from spectral fitting to the MOS1, MOS2 and pn spectra of the off-nuclear sources in NGC 7252 across the 0.5 $-$ 10.0 keV energy range (see Figure \ref{7252-blob-spec}). The models fitted include a component for Galactic absorption. The luminosities quoted are unabsorbed, for H$_{0} = 75$ km s$^{-1}$ Mpc$^{-1}$, in the energy range 0.5$-$10.0 keV. The errors quoted are the 90\% confidence limits on each single parameter of interest, and the hydrogen column density refers to the intrinsic absorption, which is fitted in addition to the fixed line-of-sight Galactic absorption.}\label{7252-blob-tab} 
\end{center}
\end{table*}

\begin{figure*}
\centerline{\epsfig{file=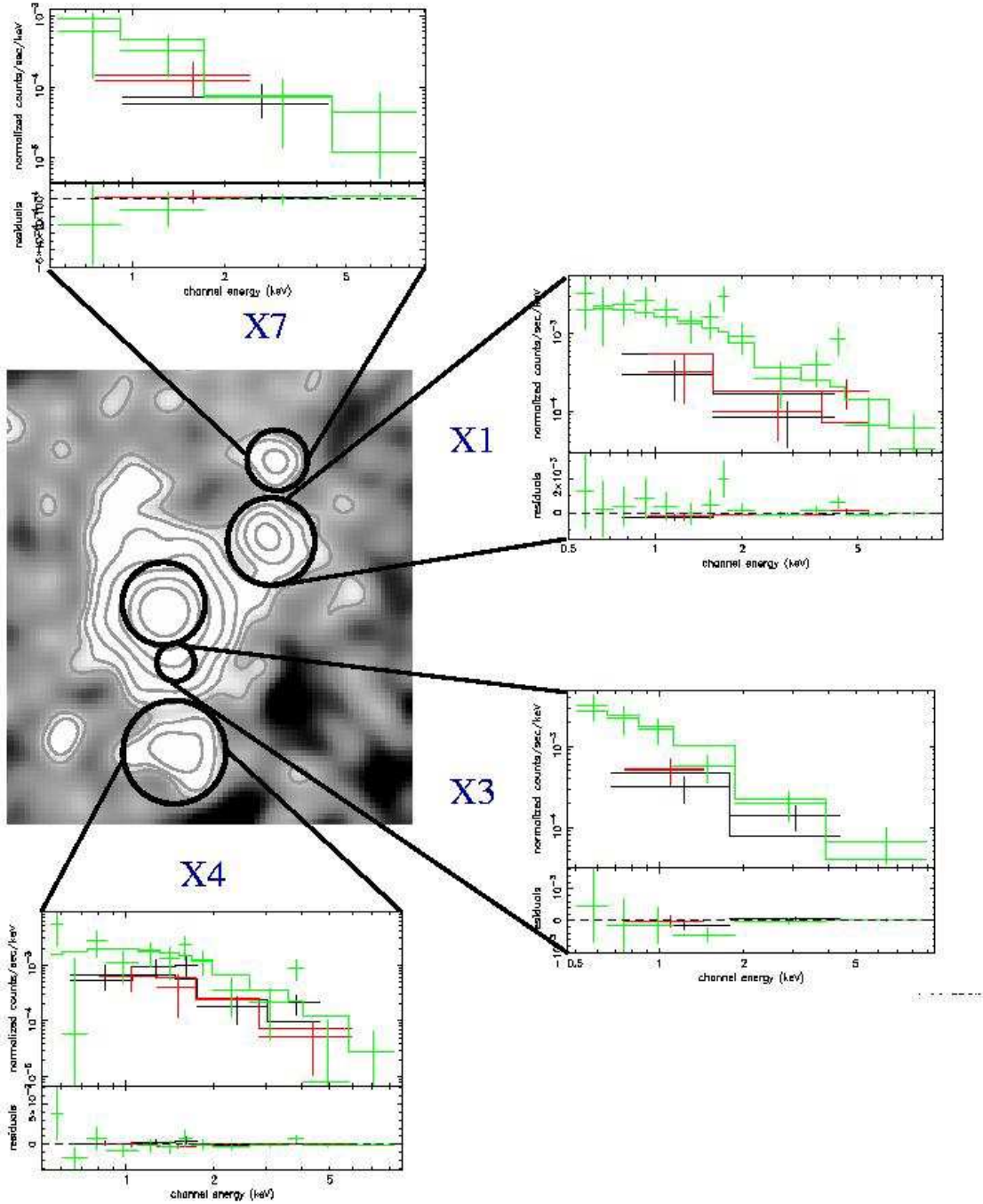,width=17cm,angle=-0,clip=}}
\caption{Spectral fits with best-fitting models as listed in Table \ref{7252-blob-tab} for the off-nuclear sources in NGC 7252. Centre: X-ray contour map (0.5$-$2.0 keV) showing the regions from which the spectra were extracted. Surrounding plots: crosses: pn (upper) and MOS (lower) spectra; solid lines: best-fitting model spectra. See Table \ref{7252-blob-tab} for details of the models and energy bands, and \S\ref{7252spectxt} for description of the fitting and discussion. The spectra in the plots have been rebinned to at least 10 counts per bin for clarity. }\label{7252-blob-spec}
\end{figure*}

We fit absorbed power-law and disk blackbody models to four of the
five off-nuclear sources detected in our \xmm\ observation of NGC
7252. The source regions are shown in Figure \ref{7252-blob-spec}. The
background regions were defined within large circles on the same CCDs
as the sources, in regions uncontaminated by detected sources. There
is likely some contamination from nuclear emission in the source
region for 7252-X3, but, from Figure \ref{7252sbr}, the surface
brightness of the nuclear region has dropped to $\sim$5\% of the peak
value for the hot diffuse gas at this distance, so the X-ray emission
will be dominated by the off-nuclear sources.

As with 3921-X4, 7252-X8 has too few counts to successfully
disentangle it from the diffuse X-ray background spectrum and
investigate its spectrum. We do not fit MeKaL models to the four
spectra, as all four of these sources are confirmed point sources (in
the case of 7252-X4$+$, potentially the superposition of two or three
point sources), and therefore do not have a hot diffuse gas
origin. The models were fitted across the 0.5$-$10.0 keV energy
range. The spectra are not grouped into bins, so the Cash statistic is
used for the fitting, as for 3921-X3. The extraction regions, observed
and best-fitting model spectra are plotted in Figure
\ref{7252-blob-spec}.

Three of these four sources (7252-X1, X3 and X7) are best-modelled by
power-law models, with indices 1.38, 1.98 and 2.06 respectively,
comparable within the uncertainties, and with no intrinsic
absorption. The fourth source, 7252-X4$+$, is better-modelled by a
disk blackbody model, with a small ($N_H$ $=$ 2.0x10$^{20}$ cm $^{-2}$)
amount of intrinsic absorption, and kT$_{inner} = $ 1.30 keV. Although
this may be contaminated with contributions from 7252-X11 and
7252-X12, the \chandra\ data shows that these two sources are less
than half as luminous as 7252-X4, so the spectrum should still be
dominated by contributions from 7252-X4. Of course the possibility
that the X-ray emission from X11 and/or X12 varies considerably over
the timescale of the two observations cannot be rejected, and one or
both of these sources may have been more more luminous in the \xmm\
data, observed four days before the \chandra\ data.

All these sources are very luminous, with L$_{X} > 10^{39}$ erg
s$^{_1}$, and therefore they are all ULXs, if they lie within NGC
7252. Indeed, they are some of the most luminous ULXs ever
observed. In \S\ref{ulxdiscuss}, we discuss the number of these
sources that we expect to be background objects.

There are too few counts to fit complex models to the \chandra\ data, so we estimate luminosities using a power-law model with an index of 1.5 and no intrinsic absorption. The \chandra-derived luminosity agrees, within the errors, with the \xmm\ luminosity for the source 7252-X7 (which is not well-constrained by the spectral fitting to the EPIC spectra). 7252-X1 and 7252-X3 have only a minor disagreement between the two observations. One might expect differences between the two estimates due to the different models used, but differences may also arise from intrinsic variation in the source. This is discussed further in \S\ref{ulxdiscuss}. The luminosity of 7252-X4, estimated from the \chandra\ observation, lies within the uncertainty of the luminosity estimated using the disk-blackbody model fitted to the \xmm\ spectrum (the best-fitting model). However, the possibility that the \xmm\ spectrum of 7252-X4$+$ is contaminated with emission from the possible \chandra\ sources 7252-X11 and 7252-X12 cannot be ruled out; the sum of the X-ray luminosities of 7252-X4, X11 and X12, estimated from the \chandra\ data, also lies within the uncertainties in the luminosity determined from the \xmm\ data.

If these low-significance sources are real, and lie within NGC 7252, they, like the more significant sources, are ULXs, with \lx\ $\ge$ 10$^{39}$ ergs$^{-1}$.

\section{Discussion}

\subsection{Starburst and dynamical ages}

The dynamical age, i.e. the age since the last peri-galactic passage that induced the merger, of NGC 3921 is inferred from its morphology. The nuclei of the progenitor galaxies have already merged, but long tails remain, and this, together with the off-centre nucleus and `sloshing' isophotes imply that the merger age is 700$\pm$300 Myr, which is a few central crossing times (t$_{cr}$ $\sim$ 5x10$^{7}$ yr, Schweizer 1996). In the case of NGC 7252, N-body simulations suggest that $\sim$ 580h$^{-2}$ Myr have elapsed since peri-galacticon, i.e. $\sim$ 1 Gyr, for H$_{0} =$ 75 km s$^{-1}$Mpc$^{-1}$ (Hibbard \& Mihos 1995). Its symmetrical isophotes and equilibrated central disk of molecular and ionised gas are consistent with it being a few 10$^{8}$ yr further evolved than NGC 3291.

The age of the starburst in NGC 3921 has been estimated by Schweizer et al. (1996) at 250$-$750 Myr, for Z= 1.0$-$0.2 \Zsolar, from colour-magnitude (V$-$I, V) diagrams of the candidate globular clusters and stellar associations, and the stellar population evolutionary synthesis models of Bruzual \& Charlot (1993). The results suggest that globular clusters have been forming for most of the time since the merger began, and possibly even prior to peri-galacticon. In NGC 7252, Miller et al. (1997), again using (V$-$I, V) colour-magnitude diagrams for globular clusters and the Bruzual \& Charlot models, estimate ages of 650$-$750 Myr for 1.0$-$0.5 \Zsolar, together with a central $<$ 10 Myr population, most likely of O stars and stellar associations.

\subsection{Ultra-luminous X-ray point sources}\label{ulxdiscuss}

In NGC 3921, excluding the LINER, we find two sources for which we
have measured intrinsic X-ray luminosities $>10^{39}$ erg s$^{-1}$,
and which, if they are indeed point-like, and reside within the
galaxy, must therefore be ULXs. 3921-X4 is most likely ultra-luminous
as well; the source-searching software assigns it $\sim$5 times as
many counts as 3921-X3 (Table \ref{3921srcs}), which has \lx\ $=$
8.32x10$^{39}$ erg s$^{-1}$. However, it is not possible to
disentangle 3921-X4 from the diffuse gas X-ray emission to perform
spectral fitting to measure its X-ray luminosity without more counts
and / or better spatial resolution.

In NGC 7252, the \xmm\ observations reveal four off-nuclear sources
with \lx\ $>$ 10$^{39}$ erg s$^{-1}$, and a fifth, unidentified
source, which is potentially a ULX as well. Source searching of the
November 2001 \chandra\ data finds five such sources with a
significance $>$ 4$\sigma$, three of which are coincident with \xmm\
sources, together with a low-significance detection of 7252-X7, also
detected by \xmm. If we include the low-significance sources, the
number of ULXs detected in NGC 7252 could be as high as
thirteen. Using the logN$-$logS relation of Tozzi et al. (2001), we
can estimate the expected number of background sources not physically
associated with the merger remnants. We expect to find 0$-$1
background sources in NGC 3921 in the \xmm\ data, and the same in NGC
7252, over the areas shown in Figures \ref{3921ebands} and
\ref{7252ebands} respectively. In the \chandra\ data, over the area
shown in Figure \ref{chandra}, $\sim$4.5$\pm$2 background sources
are expected at or above the low detection limit (2.3$\sigma$).

The difference in the number of sources detected by \xmm\ and
\chandra\ in NGC 7252 may be due in part to time variability of the
sources. The \chandra\ observation was carried out only four days
after the \xmm\ observation, so if variability does account for the
differences in the source detection and modelled luminosities, it
would have to occur on a timescale of $\sim$ a few days. Fabbiano et
al. (2003) find spectral variability in all the non-nuclear ULXs in
the Antennae, consistent with accreting compact XRB models, but this
is on timescales of two months to two years. However, the orbital
periods of HMXBs can be $\sim$1$-$10's of days, and periodic
fluctuations, which can increase the (hard) X-ray luminosity by a
factor $\sim$ 10, corresponding to the periastron passage of the
compact object, have been detected by Laycock et al. (2003), using the
Burst and Transient Source Experiment. This is consistent with our
apparent detection of variability over a timescale of 4 days.

Alternatively, the greater spatial resolution of \chandra\ may allow
detection of sources which \xmm\ can not resolve; this is almost
certainly the case with 7252-X2. The \xmm\ detection limit (for a
point source modelled by a power-law with an index of 1.5) is
$\sim$3.3x10$^{39}$ erg s$^{-1}$ in NGC 3921 and $\sim$1.7x10$^{39}$
erg s$^{-1}$ in the less-distant galaxy NGC 7252. However, the lower
significance limit, $\sigma =$2.3 represents a detection threshold,
for the same model, of 10$^{39}$ erg s$^{-1}$ in the \chandra\
observation of NGC 7252; we are not able to detect the low
significance \chandra\ sources in the \xmm\ data.

In a recent study, Swartz et al. (2003) have carried out a preliminary
analysis of 84 near-by galaxies in the \chandra\ archive, spanning the
range of Hubble morphological types, in a search for ULXs. They
detected 120 ULXs with \lx\ $>$ 10$^{39}$ erg s$^{-1}$ in these
galaxies, only nine of which have \lx\ $>$ 10$^{40}$ erg
s$^{-1}$. These reside in four galaxies, all in their `spirals \&
irregulars' class, which includes merging galaxies and
merger-remnants. {\em No} elliptical galaxies in this survey have \lx\
$>$ 5x10$^{39}$ erg s$^{-1}$. They find that both the number and mean
X-ray luminosity of ULXs increase significantly in merging and
interacting galaxies, and also correlate with far infrared (FIR)
luminosity. The number of ULXs per galaxy residing in galaxies which
are in the process of merging is $\sim$5, and the mean X-ray
luminosity is $\sim$7x10$^{39}$ erg s$^{-1}$. Although the source
detection thresholds in the heterogeneous \chandra\ study and our
\xmm\ and \chandra\ data are different, so we cannot make a
statistical comparison, the numbers and mean luminosities of the ULX
candidates we find in NGC 3921 and NGC 7252 do not conflict with the
preliminary results of Swartz et al.. They do, however, point out that
the trend towards brighter ULXs in galaxies with higher numbers of
ULXs may be a statistical fluctuation, and is quite weak.

NGC 3921 and NGC 7252, therefore, contain some ultra-luminous X-ray sources. Most of those in NGC 7252 are confirmed as point sources by the \chandra\ observation (Figure \ref{chandra}, Table \ref{chandrasrcs}). The spatial resolution of \xmm\ is not sufficient to confirm the point-like nature of 3921-X2, but this is the most luminous X-ray source detected in these observations. We can compare our results with the \chandra\ ACIS-S observations of the merging system NGC 4038/4039 (the Antennae galaxies, Zezas et al. 2002). Eight point sources with \lx\ $>$ 10$^{39}$ erg s$^{-1}$ are detected in this system (assuming a 5 keV bremsstrahlung model and Galactic line-of-sight absorption). The brightest of these sources has a luminosity of 8.9x10$^{39}$ erg s$^{-1}$ (corrected to H$_{0} =$ 75 km s$^{-1}$Mpc$^{-1}$), measured in the 0.1$-$10.0 keV range. The brightest off-nuclear sources in our observations are 3921-X2 and 3921-X3, which have X-ray luminosities \lx\ $=$ 1.38x10$^{40}$ and 1.05x10$^{40}$ erg s$^{-1}$ respectively. So, even in the narrower range (0.5$-$10.0 keV) across which we do our spectral modelling, these two sources are brighter than the brightest Antennae source. However, Miller et al. (2003) fit more complex, disk blackbody plus absorbed power-law models to the \xmm\ spectra of a sub-set of four of these eight bright Antennae sources, and determine X-ray luminosities of 1.0$-$3.7x10$^{40}$ erg s$^{-1}$ for them, in the energy range 0.3$-$10.0 keV. 

The best-fitting power-law models for the eight Antennae sources have indices ranging from 1.21 to 1.94, with intrinsic $N_H$ $=$ 5.0$-$2.6 x 10$^{20}$ cm$^{-2}$ (Zezas et al. 2002), which is comparable  with what we find in NGC 3921 and NGC 7252 for the sources best-fit by power-law models.

\chandra\ observations of the interacting galaxy system NGC 4676 (the
Mice), find five ULXs with \lx\ \gs5x10$^{39}$ (Read 2003). This
system is similar to, though less evolved than, the Antennae
galaxies. Although in this more distant galaxy (D $=$ 88 Mpc) source
confusion is likely even with \chandra 's spatial resolution, this is
again consistent with the picture of enhanced numbers of very luminous
XRBs in merging systems.

3921-X2 and 7252-X4$+$ are best-modelled by disk blackbody spectra,
with T$_{inner} =$ 0.81 and 1.30 keV respectively; these sources are
plausibly black hole systems in the high/soft state. For these models,
kT $\sim$ 1 keV(M/\Msolar)$^{-1/4}$ (e.g. Makishima et al. 2000), so
these temperatures are consistent with the presence of stellar-mass
black holes in these objects ($\sim$ 23.3\Msolar\ and 3.5 \Msolar,
respectively). However, it should be noted, as pointed out in
\S\ref{7252onsrcs}, that the \xmm\ spectrum of 7252-X4$+$ could be
contaminated with emission from the sources 7252-X11 and X12, although
these sources are only detected with low significance ($\sim$
3$\sigma$). Note also that the other spectral models are not ruled out
by the fitting statistics.

Observations of early-type galaxies suggest that 40$-$70\% of their ULXs 
are associated with young globular clusters (Angelini et al. 2001;
Maccarone et al. 2003). Therefore, we have used existing HST optical
observations of globular clusters (NGC 3921: Schweizer et al. 1996;
NGC 7252: Miller et al. 1997) to look for the optical counterparts to
the X-ray sources in these two galaxies.

If the \xmm\ coordinates (absolute astrometry $\sim$4$-$5\arcsec) of
NGC 3921 are accepted without any further correction, then the
position of the X-ray nucleus is within 2\arcsec\ of the HST-optical
nucleus, and 3921-X2 corresponds to the stellar associations 127
and/or 128 from Schweizer et al. (1996). The coordinates of these
associations are listed in Table \ref{3921pos}. There is no
coincidence between 3921-X3 and any optical source, and the nearest
optical sources to 3921-X4 are a candidate globular cluster (GC 68)
and stellar association (A 126), both lying $>$ 5\arcsec\ from
3921-X4. However, at the distance of NGC 3921, stellar associations
have effective radii which are typically $\la$ 1\arcsec, and for
globular clusters, it is even less, \ls\ 0.03\arcsec. The astrometric accuracy of HST is $\sim 1-3$\arcsec\ (e.g. Brammer et al. 2002).

If we accept the \chandra\ coordinates (absolute astrometry
$<$1\arcsec) for NGC 7252 without any further corrections, the X-ray
and HST-optical positions of the nucleus agree to
$\Delta$RA/$\Delta$Dec = -0.3\arcsec/$+$0.4\arcsec, and there is good
agreement between the positions of 7252-X1, X3 and X7 in the \xmm\ and
\chandra\ data sets (Figure \ref{chandra}). The \xmm\ detection of
7252-X4$+$ agrees less well with the position of the \chandra\
detection, but this is probably due to confusion between the sources
7252-X4, X11 and X12. The good agreement means we can be confident in
comparing the X-ray source positions with optical
observations. 7252-X2 corresponds closely (within 2\arcsec) to both
the stellar association MSW97-45 and the globular cluster MSW97-9
(Miller et al. 1997). The other sources have no such optical
coincidences (except for 7252-X5 which is a foreground star). The
absolute astrometry of \chandra\ is $<$1\arcsec.

Hence, for these two post-merger galaxies, the fraction of candidate
ULXs associated with globular clusters / stellar associations is less
than 20\% (assuming all these sources are actually associated with the
merger remnants).

If XRBs are created in globular clusters (White
et al. 2002, Maccarone et al. 2003), then it is possible that they could be ejected from their parent clusters. 3-body interactions are likely to be the dominant mechanism for this ejection. These could almost always eject XRBs with M$_{BH}$ $<$ 10
\Msolar\ from globular clusters, and virtually inevitably from young
stellar clusters, which have lower escape velocities (Kulkarni, Hut \&
McMillan 1993; Sigurdsson \& Hernquist 1993). From Quinlan (1996), the
velocity of the kick applied to a binary in a 3-body interaction,
$v_{kick}$, is:

$$v_{kick}/km s^{-1} \sim 40 (m/10 \Msolar)^{3/2}(M/100 \Msolar)^{-1},$$

where $m$ is the mass of the companion star, and $M$ the mass of the
compact object. So, for the $\sim$ 10 kms$^{-1}$ required to escape
from a young stellar cluster (Coleman Miller \& Colbert 2001), then if
$m\ \sim\ 20$ \Msolar, $M$ can be as high as 10$^{3}$
\Msolar. However, for a dense globular cluster, where $v_{kick}$ \gs
30 kms$^{-1}$ (Webbink 1985), then even for $m = 20$, $M < 400$ \Msolar.

It is also possible that XRBs may be `kicked' out of
their parent globular cluster during the collapse of the compact
object by which they are powered. Even at very modest velocities
(v$\sim$30$-$100 km s$^{-1}$, Cordes \& Chernoff 1998; Fryer \&
Kalogera 2001, respectively), they could travel several hundred
parsecs over the timescales suggested above (0.3$-$9 kpc in 10$-$300
Myr at 30 km s$^{-1}$). The distance of 3921-X4 from its nearest
optical neighbours is $\sim$1.9 kpc, so it is plausible that it has
been ejected from one of GC 68 or A 126. 3921-X2 and 7252-X2 are the
only sources which correspond to one (or two) stellar
associations. The `kick' velocity from supernova collapse is inversely
proportional to the mass of the system (Cordes \& Chernoff 1998). This
suggests that ULXs residing at any significant distance outside the dense regions in which they are
most likely formed must either be stellar-mass black holes, or must not have been ejected via the kick from supernova collapse. In the supernova kick scenario, therefore, 3921-X2 and 7252-X2, which lie closest to their putative parent globular clusters,
are the most likely candidates for intermediate-mass black holes
(IMBHs), with masses 10$^{2}$ \ls\ M/\Msolar\ \ls\ 10$^{4}$, which
could reach ULX luminosities with isotropical sub-Eddington
emission. 

It is difficult, therefore, to distinguish between an isotropic IMBH
or beamed HMXB origin for ULXs on basis of estimating mass from
positional considerations. Rigorous constraints on mass are needed
from observations of the radial velocities of optical counterparts, or
gravitational waves. 

IMBHs could form either via multiple mergers/accretion, or from the
collapse of massive, negligible metallicity, Population III stars. Both
these scenarios require a high stellar density for there to be a
reasonable probability for interactions to occur, either for the
multiple merging / accretion which forms an IMBH in the first case, or
for the capture of a massive stellar companion in the second. Their
existence in ULXs was invoked in order to avoid the necessity for
super-Eddington emission from ULXs. However, King (2003), using
\chandra\ observations of the Cartwheel galaxy, suggests that, even for
IMBHs accreting from massive stars, the inferred production rate and
low accretion efficiency would require super-Eddington accretion. He
postulates that, unless IMBHs accrete from an unknown, non-stellar
source, with very specific properties, most, if not all, ULXs
associated with star-formation must be HMXBs. The case in favour of an
IMHB origin for ULXs is reviewed in detail in Coleman Miller \&
Colbert (2003).

It is possible that some of the detected X-ray sources are in fact
groups of less-luminous X-ray sources. The spatial resolution of
\chandra\ ($\sim$0.5\arcsec) corresponds to $\sim$150 pc at the
distance of NGC 7252 (63.0 Mpc). Unbound stellar associations have
radii of order twice this (Blauw 1964), while globular clusters, which
are gravitationally bound, have half light radii of \ls 10 pc
(Schweizer et al. 1996), much less than the resolution at this
distance. 7252-X2, the only off-nuclear X-ray source in the \chandra\
data with a likely optical counterpart, could therefore be the sum of
X-ray emission from multiple objects residing in a globular cluster or
stellar association. 

The contribution of X-ray emission from normal O stars can be readily discounted. A typical solar metallicity O-type star emits
$\sim$3.5x10$^{34}$ ergs$^{-1}$ in the 2$-$10 keV range (Helfand \&
Moran 2001), so it would require $\sim$6x10$^{4}$ O-type stars to
result in the 2$-$10 keV luminosity seen in this object. This is $\sim$3 times
the total number of O-type stars in the entire Milky Way, and they would all
have to reside in a region of radius $\sim$150 pc. The lower spatial resolution of \xmm\ corresponds to $\sim$ 2kpc, but the observed luminosities of the off-nuclear sources detected with \xmm\ in NGC 3921 could only be reached with $\sim$ a few x10$^{5}$
O-type stars.

A more promising possibility for source confusion is a superposition of bright X-ray binaries. We can estimate the number of 10$^{38}-$ 10$^{39}$ ergs$^{-1}$ sources we would expect to find in NGC 3921 and NGC 7252 by comparison with the well-studied Antennae galaxies. Fabbiano, Zezas \& Murray (2001) find 23 sources in this energy range in the Antennae. Normalising this number by the B$-$band luminosity, which is well-correlated with the total point source luminosity (Colbert et al. 2003), and by comparison with the X-ray luminosity function in the Antennae galaxies, we expect $\sim$26 and 29 such sources in NGC 3921 and NGC 7252 respectively, with total aggregate luminosities of $\sim$4.2 and 4.9x10$^{39}$ ergs$^{-1}$. 

One could imagine a large group of XRB sources associated with an exceptional star cluster, formed in the extreme conditions of a galaxy merger. Of our ULX sources, only 3921-X2 and 7252-X2 have potential cluster counterparts. 3921-X2 has more than twice the X-ray luminosity estimated above for {\it all} the 10$^{38}-$ 10$^{39}$ ergs$^{-1}$ sources in NGC 3921, and even for the lower luminosity 7252-X2, the majority of the expected bright XRB sources in NGC 7252 would have to reside in a single region \ls 150 pc across. If the proposed optical counterparts to 3921-X2 or 7252-X2 were truly exceptional, for example, if they hosted massive star-formation, then the high density of 10$^{38}-$ 10$^{39}$ ergs$^{-1}$ sources might be conceivable, but this is not the case (Miller et al. 1997; Schweizer et al. 1996). 

It seems most likely, therefore, that the candidate ULXs we observe in NGC 3921 and NGC 7252 are mostly single ultra-luminous objects. They are 
likely to be HMXBs, which are associated with recent
star-formation. We see more ULXs, with higher luminosities, than are
observed in galaxies without recent star formation (e.g. Swartz et
al. 2003; Irwin et al. 2003). O$-$B stars at the end of their main
sequence lifetime are the companions to the compact objects (neutron
stars or black holes) in HMXBs, and take $\sim$10$-$300 Myr to reach
this point in their evolution. This timescale is entirely plausible,
given on-going star-formation since peri-galacticon in both these
galaxies.

\begin{table}
\begin{center}
\begin{tabular}{ccccc}

\hline     
	Object	&  RA(J2000) /  &    DEC(J2000) /    &   $\sim V$   &   $V-I$       \\
	      	&       h:m:s   &         d:m:s      &       &             \\
\hline
	3921-X2	& 11:51:07.63 &  +55:04:16.53 &       &             \\
	A 127	& 11:51:07.67 &  +55:04:17.8  & 24.9 & +0.05$\pm$0.18  \\ 
	A 128   & 11:51:07.72 &  +55:04:16.3  & 25.2 & +0.28$\pm$0.21  \\ 
\hline
\end{tabular}

\caption{Stellar associations corresponding to 3921-X2 (Schweizer et al. 1996).}\label{3921pos}
\end{center}
\end{table}

\subsection{X-ray halo regeneration scenarios}

Read and Ponman (1998), in a study of the X-ray evolution of on-going
mergers using ROSAT data, find that the late stages of mergers are
under-luminous in X-rays compared with typical ellipticals. Although
when the two nuclei coalesce, massive extensions of hot gas are
observed (up to 10$^{10}$ \Msolar, e.g. in the Antennae, Arp 220, NGC 2623),
after this peak, the X-ray luminosity decreases. In fact, in
post-merger remnants, \lx\ / L$_{B}$ is $\sim$ 5 times less than that
of a typical elliptical (e.g. Fabbiano et al. 1992). If major mergers
are indeed the progenitors of normal elliptical galaxies, the X-ray
halo must be in some way regenerated following the merging incident.

In this section, we use our \xmm\ observations to investigate possible
mechanisms for the regeneration of X-ray haloes in NGC 3921 and NGC
7252. As we have outlined in the introduction, there are three hypotheses proposed for rebuilding the hot halo: the infall of tidally stripped cold gas, mass loss from strong stellar winds, and the return of hot gas expelled earlier in the merging process.

\subsubsection{Infall of tidally stripped cold gas}

The HI gas in the two merger remnants is clearly associated with the
debris from their gas-rich parent-galaxies, i.e. the tails and western
loop in NGC 7252 and the southern tail in NGC 3921. HI gas is not,
however, seen to the north of NGC 3921 (M$_{HI}$/L$_{R}<$ 0.02
\Msolar/\Lsolar, Hibbard \& van Gorkom 1996). 

In NGC 7252, the HI gas
is not obviously associated with X-ray emission. In NGC 3921, the
picture is somewhat different. The extremely luminous source to the
south of the nucleus (3921-X2) is clearly associated with the HI tail,
H$\alpha$ emission and a large group of star clusters, suggesting that
it is related to recent star formation.

The spatial resolution of \xmm\ is not sufficient to determine whether this source is point-like, and the possibility that this source is not an X-ray binary but a local region of shock-heated diffuse gas cannot be dismissed on the basis of the spatial data. However, it is unlikely that the region is being shock-heated via winds from close-by stellar clusters; although these can reach temperatures $>$ 2 keV (the temperature of the best fitting MeKaL model to 3921-X2), their X-ray luminosities are several orders of magnitude lower ($<$ 10$^{36}$ ergs$^{-1}$) than in 3921-X2 (e.g. Stevens \& Hartwell 2003). 

The possibility that 3921-X2 is a shock region due to the infall of cold gas still remains. However, we can estimate the upper limit to the post-shock temperature, and compare
this with the temperature of the best-fitting MeKaL model of this
source. From Cavaliere et al. (1997):
\begin{equation}
 kT_{2} = \frac{1}{3}m_{p}\nu ^{2} + \frac{7}{8}kT_{1}, 
\end{equation}
where kT$_{1}$ is the pre-shock temperature, kT$_{2}$ is the
post-shock temperature, m$_{p}$ is the proton mass, and $\nu$ is the
pre-shock velocity of the infalling cold gas. The upper limit of the
gas velocity is the velocity it would gain from falling into the
galaxy from rest at infinity. We take the dynamical mass of the galaxy
to be 1.06x10$^{11}$\Msolar\ (Hibbard \& van Gorkom 1996), and assume
the pre-shock cold gas temperature is negligible. Putting these
assumptions into equation 2, we estimate the upper limit of the
post-shock temperature to be $\sim$0.96 keV ($\sim$10$^{7}$K). This is
well below the lower limit of the best-fitting MeKaL model (1.94 keV),
and suggests that shocked gas is not the origin of 3921-X2. 

However, it should be noted that freshly shocked gas may not be in either thermal or  ionization equilibrium, and this can alter the observed properties of the post-shock gas. Takizawa (1999) discusses the thermal equilibrium timescale, noting that the ion temperature is initially much higher than the electron temperature in the post-shock region (since most of the incoming kinetic energy is in the ions), and that the two temperatures equilibrate on an electron-ion interaction timescale. However, this effect would lead to a {\it low} observed electron bremsstrahlung temperature, whereas we see a best fitting MeKaL temperature {\it higher} than the upper limit to the post-shock temperature. Moreover, the equilibration timescale at the temperature and gas density in the region of 3921-X2 ($\sim$0.4 Myr) is much shorter than the post-shock timescale, inferred from the time elapsed since star formation (several hundred Myr).

Calculations of the thermal and dynamical evolution of a fast adiabatically expanding gas (Breitschwerdt \& Schmutzler 1999) show that a delay in recombination can lead to departures from ionization equilibrium. A full set of dynamically and thermally self-consistent equations is needed to calculate the timescale for recombination, which is beyond the scope of this paper. However, Breitschwerdt \& Schmutzler suggest that the dynamical timescale for a hot plasma (T$\ge 10^{6}$K) is much shorter than the intrinsic timescales (e.g. for recombination, ionization etc.). The ionization non-equilibrium gives rise to the appearance of emission lines representative of higher temperatures than the actual gas temperature, and the over-estimation of the temperature if the spectrum is fitted with collisional ionization equilibrium models. However, for 3921-X2, the upper limit to the post-shock temperature is  $\sim$0.96 keV, and it is difficult to see how a temperature as high as the best fitting MeKaL temperature, 2.18 keV, could be mimicked by this process. In addition, we do not see any emission lines in the spectra (Figure \ref{3921-blob-spec}) which we would expect from this scenario. This lack of emission lines is consistent with the very low abundance of the best fitting MeKaL model. Hence, we conclude that the high temperature of the best fitting MeKaL model is inconsistent with a shock-heated gas model.

In NGC 7252, we do not see an X-ray emission region analogous to that
of 3921-X2, i.e. with associated clusters, HI and H$\alpha$
emission. However, in Figure \ref{7252Hregs}, we see a `cavity' in the
0.5$-$2.0 keV X-ray emission to the north, which appears to correspond
to the north-western HI tail where it crosses the hot gas. Since the
tail comes across the {\it front} of the remnant body (Schweizer 1982;
Hibbard \& van Gorkom 1996), the increased HI surface density in the
cavity region (15$-$20 x10$^{19}$ cm$^{-2}$, figs. 5c and 5d in
Hibbard et al. 1994) strongly suggests that the `cavity' is due to
X-ray absorption by the neutral gas of the tidal tail in the
foreground. This effect may of course be enhanced by the positions of
7252-X1 and 7252-X7.

If X-ray halos are indeed regenerated by infall along tidal tails, then maybe insufficient time since peri-galacticon has passed for substantial regeneration to have occurred (e.g. Hibbard \& Mihos 1995). Considerable HI is evident out at least as far as the optical tails in NGC 7252 and beyond the optical loops to the south of NGC 3921. Hibbard \& Mihos (1995) model the kinematics of the HI gas in NGC 7252 using N-body simulations. They calculate that, of the HI originally residing in the spiral progenitors of this object, at least 3x10$^{9}h^{-2}$\Msolar\ has already fallen inwards. However, the central regions of NGC 7252 are HI-poor, so atomic gas falling into the remnant body from the tails must be converted to other forms, at a rate of 1$-$2 \Msolar\ yr$^{-1}$. The process of infall is a natural explanation for the conversion of HI to hot gas, stars and cold molecular gas. This hypothesis is supported by observations of both NGC 3921 and NGC 7252 which show the presence of all these three phases in the remnant bodies (e.g. Whitmore et al. 1993; Hibbard et al. 1994; Hibbard \& van Gorkum 1996; Schweizer 1996; Schweizer et al. 1996; Yun \& Hibbard 2001).

Using the results of the Hibbard \& Mihos simulations, we can estimate if there is sufficient HI still to fall back into the remnant body of NGC 7252 from the tidal tails, to increase the hot gas mass (and hence luminosity) to that seen in a typical elliptical galaxy.

The simulations indicate that $\sim$6x10$^{8}h^{-2}$\Msolar\ of HI will return to within 5R$_{e}$ over the next 3$h^{-2}$ Gyr, i.e. one fifth as much again as has already fallen into the central regions. We estimate the mass of hot gas already present in the remnant body from the parameters of the MeKaL components fitted to the nuclear region, and assuming spherical symmetry. The mass is $\sim$3x10$^{8}$\Msolar, within the spectral extraction region. `Typical' X-ray bright elliptical galaxies (not including brightest group and massive cluster ellipticals, which may be considered a different class) have hot gas masses of 10$^{9}-$10$^{10}$\Msolar\ (e.g. Forman et al. 1985). Even if {\em all} the HI gas still to return to the remnant body is converted to hot gas, with none involved in star formation or conversion to molecular gas, the total hot gas mass would be just below 10$^{9}$\Msolar, and it is unlikely that the conversion to hot gas is anywhere near this efficient. For comparison,  $\sim$3x10$^{9}$\Msolar\ of HI has already fallen back into the body of the galaxy, but we see only $\sim$3x10$^{8}$\Msolar\ of hot gas, which, assuming all the hot gas has been converted from HI, with no contributions from stellar mass loss, represents an efficiency of only 10\%. It is hard to see that HI infall in this galaxy would create sufficient hot gas to reach the mass seen in typical elliptical galaxies.

We know that the northern tail of NGC 3291 is HI-poor, but the southern tail contains $\sim$ 2x10$^{9}h^{-2}$\Msolar\ of HI, with the tail becoming more gas-rich along its length (Hibbard \& van Gorkom 1996). However, there are no simulations of the kinematics of the atomic gas in NGC 3921 in the literature, so we do not know what proportion of this gas will return to the remnant body. In addition, the irregular X-ray structure of NGC 3921 invalidates the assumption of spherical symmetry used to calculate the hot gas mass in NGC 7252, so we cannot make the same calculation. However, assuming its similar X-ray luminosity to NGC 7252 indicates a similar gas mass, this galaxy would require an additional mass of hot gas \gs5x10$^{8}$\Msolar\ to take it to typical values. This only seems possible if the HI kinematics and/or conversion to hot gas in NGC 3921 are quite different from those in NGC 7252.

Given the differences between the predicted hot gas masses of typical ellipticals and those predicted from our calculations involving HI masses, it is difficult to believe that HI infall can be solely responsible for the regeneration of the X-ray halo to luminosities typical of elliptical galaxies. Additional support for the existence of another mechanism for halo regeneration comes from the work of O'Sullivan et al. (2001a). They see a long-term trend of increasing X-ray luminosity with mean stellar age, extending over 15 Gyr, long after the HI tails seen in mergers have been completely assimilated into the body of the galaxy. If the X-ray luminosity continues to rise after the completion of HI infall, then, even if HI conversion is responsible for the presence of some of the hot gas, there must be an additional source, which contributes to the hot gas mass after virialisation, for example, stellar mass loss, as discussed below.

\subsubsection{Stellar winds}

The models of Ciotti et al. (1991) investigate whether mass loss from stars could regenerate the X-ray halo. They predict a super-sonic wind-driven flow in the early stages of post-merger elliptical evolution, arising from two sources: mass-loss from evolving stars, and secularly declining heating from Type I supernovae. During this stage, a galaxy is X-ray faint, and the amount of hot X-ray gas, and hence the X-ray luminosity, decreases. This stage can persist for $\sim$ 2 Gyr, and is followed by a sub-sonic outflow. As the flow velocity decreases, this flow allows the build-up of hot gas, and results in rising X-ray luminosity over timescales \gs 10 Gyr. In this model, NGC 3921 and NGC 7252 would be in the first, wind-driven stage, and indeed, they are X-ray faint. The gas density and temperature are expected to fall off smoothly with radius, without the presence of shocks. We see no evidence that this is not the case (if we are correct in assuming that 3921-X2 is a ULX). However, in order to model the behaviour of gas density and temperature with radius in detail, we would need both better spatial resolution than \xmm\ can offer, and more counts. The sub-sonic outflow stage of this model, where X-ray luminosity increases gradually with time, is consistent with the observations of elliptical galaxies discussed by O'Sullivan et al. (2001a).

\subsubsection{Infall from an outer reservoir of hot gas}

The third possible mechanism for the regeneration of the X-ray halo is
the return of the hot gas which was expelled from the main body of the
galaxy during the nuclear merging stage. This expelled gas can reach
large radii before stalling. It then falls back into the galaxy, and
shocks are expected as it collides with the existing halo gas. These
outflows are typically asymmetrical (e.g. Read \& Ponman 1998). We see
little evidence for the regeneration of the X-ray halo via this
mechanism in NGC 3921 and NGC 7252. We see little X-ray emission at
large radii in either galaxy. This does not mean that we can
reject the possibility that there is significant hot gas outside the
main remnant bodies which is as yet too diffuse to have been detected
in these observations. However, although the infall of hot gas may be delayed with respect to the cold gas, as the hot gas reaches larger radii than the cold, it will still occur on a galaxy dynamical time-scale (O'Sullivan et al. 2001b). As with the infall of cold gas, it is hard to reconcile this with the long-term (\gs 10Gyr) trend of increasing X-ray luminosity with mean stellar age observed by O'Sullivan et al. (2001a; 2001b).

\section{Conclusions}

We summarise our conclusions as follows:

\vspace*{0.34cm}

\noindent
{\bf The ULX populations}

\vspace*{0.15cm}

    In both NGC 3921 and NGC 7252, we observe ultra-luminous non-nuclear X-ray point sources. Where spectral fitting has been possible, the spectra are well-fitted by either power-law models (3921-X3, 7252-X1, 7252-X3, 7252-X7) or a disk blackbody accretion disk model (3921-X2, 7252-X4$+$). These energy distributions are typical of XRBs, and we suggest that these sources are HMXBs, associated with the recent star formation in these two galaxies. However, the possibility that the three off-nuclear sources in NGC 3921 are not point-like cannot be ruled out at present. The point-like nature of the four \xmm\ sources in NGC 7252 is confirmed by \chandra\ observations, which also reveal a further two XPSs, so that the total number of confirmed X-ray point sources observed in NGC 7252 (assuming all are associated with the remnant) is six. 7252-X8 has no corresponding \chandra\ detection, and hence cannot be confirmed as a point source. The differences in the evolutionary stages and progenitor galaxies of NGC 3921 and NGC 7252 do not appear to be reflected in their ULX populations. The numbers and luminosities of the candidate ULXs observed in NGC 3921 and NGC 7252 are compatible with the conclusion of Swartz et al. (2003), that more luminous ULXs are found in interacting and merging galaxies than in other morphological types. 

\vspace*{0.34cm}

\noindent
{\bf The distribution of hot diffuse gas}

\vspace*{0.15cm}

In NGC 7252, the more evolved merger remnant, the diffuse X-ray gas is
more relaxed: it is centred on the optical centre, is more symmetrical
in the central regions and is cooler than the hot diffuse gas in NGC
3921. This is most likely a result of its greater dynamical age. Hot
gas in NGC 7252 has a more compact core than that in NGC 3921,
reflecting differences between the progenitors of these two galaxies
and their merger dynamics. When compared to NGC 3921, the more violent
star formation in NGC 7252 is not reflected in a higher X-ray
luminosity for the hot diffuse gas.

\vspace*{0.34cm}

\noindent
{\bf X-ray halo regeneration} 

\vspace*{0.15cm}

Although substantial masses of HI are available to fall into the
remnant bodies over the next several Gyr, it seems unlikely that this
is sufficient to increase the mass of hot gas in either NGC 3921 or
NGC 7252 to the masses (and hence luminosities) seen in typical
ellipticals. We see no compelling evidence that massive X-ray halo
regeneration is occurring, even in these late-stage merger
remnants. It is likely that regeneration occurs over much longer
timescales, of up to 15 Gyr, predominantly via mass loss from evolving stars, in a sub-sonic outflow stage commencing $\sim$ 2 Gyr after the merging event (Ciotti et al. 1991).

\vspace*{1.0cm}

\noindent 
{\bf ACKNOWLEDGEMENTS}\\
   \\

We would like to thank the referee, Andreas Zezas, for his helpful comments. We acknowledge use of The Chandra Data Archive (CDA), part of the Chandra X-Ray Observatory Science Center (CXC), which is operated for NASA by the Smithsonian Astrophysical Observatory. The Digitized Sky Survey was produced at the Space Telescope Science Institute under U.S. Government grant NAG W-2166. The images of these surveys are based on photographic data obtained using the Oschin Schmidt Telescope on Palomar Mountain and the UK Schmidt Telescope. The plates were processed into the present compressed digital form with the permission of these institutions. LAN, TJP and AMR acknowledge the support of PPARC; LAN and AMR through the award of PDRAs and TJP through a PPARC Senior Fellowship. FS gratefully acknowledges partial support from the National Science Foundation through grant AST-0205994. 
   \\
   \\
   \\
   \\
   \\
{\bf REFERENCES}\\
    \\
Awaki H., Matsumoto H.,  Tomida H., 2002, ApJ, 567, 892 \\
Angelini L., Loewenstein M., Mushotsky R.F., 2001, ApJ,557L, 35\\
Belczynski K., Kalogera V., Zezas A., Fabbiano G., 2003, submitted to ApJ, astro-ph/0310200\\
Blauw A., 1964, ARA\&A, 2, 213\\
Brammer G., Whitmore B.C., Koekemoer A.M., 2002, in The 2002 HST Calibration Workshop : Hubble after the Installation of the ACS and the NICMOS Cooling System, Proceedings of a Workshop held at the Space Telescope Science Institute, Baltimore, Maryland. Ed. Arribas S., Koekemoer A., Whitmore B.C.. Baltimore, MD: Space Telescope Science Institute, 2002., p.331\\
Breitshwerdt D., Schmutzler T., 1999, A\&A, 347, 650\\ 
Bruzual G.A., Charlot S., 1993, ApJ, 405, 538\\
Cavaliere A., Menci N., Tozzi P, 1997, ApJ, 484L, 21 \\
Ciotti L., Pellegrini S., Renzini A., D'Ercole A., 1991, ApJ, 376, 380\\
Colbert E.J.M., Mushotsky R.F., 1999, ApJ, 519, 89\\
Colbert E.J.M., Heckman T.M., Ptak A.F., Strickland D.K., 2004, astro-ph/0305476\\
Coleman Miller, M., Colbert E.J.M., 2003, invited review submitted to the International Journal of Modern Physics, astro-ph/0308402\\
Cordes J.M., Chernoff D.F., 1998, ApJ, 505, 315\\
Dickey J.M. \& Lockman F.J., 1990, ARA\&A, 28, 215\\
Deeg et al., 1998, A\&A, 129, 455 \\
Ebisuzaki T., Makino J., Tsuru T.G., Funato Y., Portegies Zwart S., Hut P., McMillan S., Matsushita S., Matsumoto H., Kawabe R., 2001, ApJ, 562, L19\\
Fabbiano G., 1995, AAS, 186, 4601\\
Fabbiano G., Kim D.-W., Trinchieri G., 1992, ApJS, 80, 531 \\
Fabbiano G., Zezas A. \& Murray S.S., 2001, ApJ, 554, 1035\\
Fabbiano G., Zezas A., King A.R., Ponman T.J., Rots A., Schweizer F., 2003, ApJ, 584, L5\\
Faber S. M., Tremaine S., Ajhar E.A., Byun Y-I., Dressler A., Gebhardt K., Grillmair C., Kormendy J., Lauer T.R., Richstone D., 1997, AJ, 114, 1365\\
Forman W., Jones C., Tucker W., 1985, ApJ, 293, 102\\
Fryer C.L., Kalogera V., 2001, ApJ, 554, 548\\
Genzel R., Tacconi, L.J., Rigopoulou D., Lutz, D., Tecza M., 2001, ApJ, 563, 527\\
Helfand \& Moran, 2001, ApJ, 554, 27\\
Hibbard J.E., Guhathakurta P, van Gorkom J.H., Schweizer F., 1994, AJ, 107, 67\\
Hibbard J.E., van der Hulst J.M., Barnes J.E., Rich R.M., 2001, AJ, 122, 2969\\
Hibbard J.E., van Gorkom J.H., 1996, AJ, 111, 655\\
Hibbard J.E., Mihos, J.C., 1995, AJ, 110, 140\\
Irwin J., Sarazin C., 1998, ApJ, 499, 650\\
Irwin J.A., Bregman J.N., Athey A.E., 2003, accepted by ApJ Letters, astro-ph/0312393\\
Kaastra, J.S. 1992, An X-Ray Spectral Code for Optically Thin Plasmas (Internal SRON-Leiden Report, updated version 2.0) \\
King A.R., 2002, MNRAS, 335L, 13\\
King A.R., 2003, accepted by MNRAS, astro-ph/0309450\\
King A.R., Davies M.B., Ward M.J., Fabbiano G., Elvis M., 2001, ApJL, 552, 109\\
Kulkarni S.R., Hut P., McMillan S., 1993, Nature, 364, 421\\
Laycock S., Corbet R.H.D., Coe M.J., Marshall F.E., Markwardt C., Edge W., 2003, MNRAS 338, 211 \\
Lee H.M., 1995, MNRAS, 272, 605\\
Lee M.H., 1993, ApJ, 418, 147\\
Liedahl D.A., Osterheld A.L., Goldstein W.H. 1995, ApJL, 438, 115 \\
Maccarone T.J., Kundu A., Zepf S.E., 2003, ApJ, 586, 814\\
Maddox, Sutherland, Efstathiou, Loveday, 1990 MNRAS, 243,692\\
Makishima K., Maejima Y., Mitsuda K., Bradt H.V., Remillard R.A., Tuohy I.R., Hoshi R.,  Nakagawa M., 1986, ApJ, 308, 635 \\
Makishima K. et al. 2000, ApJ, 535, 632\\
Matsushita K. et al., 1994, ApJ, 436L, 41\\
Mewe R., Gronenschild E.H.B.M., van den Oord G.H.J. 1985, A\&AS, 62, 197 \\
Mewe R., Lemen J.R., van den Oord G.H.J. 1986, A\&AS, 65, 511 \\
Miller B.W., Whitmore B.C., Schweizer F., Fall M.S., 1997, AJ, 114, 2381\\
Miller J.M., Zezas A., Fabbiano G., Schweizer F., 2003, submitted to ApJ, astro-ph/0305488\\ 
Mitsuda K., Inoue H., Koyama K., Makishima K., Matsuoka M., Ogawara Y., Suzuki K., Tanaka Y., Shibazaki N., Hirano T., 1994, PASJ, 36, 741 \\
O'Sullivan E.J., Forbes D.A., Ponman T.J., 2001a, MNRAS, 328, 462\\
O'Sullivan E.J., Forbes D.A., Ponman T.J., 2001b, MNRAS, 324, 420\\
Quinlan G.D., 1996, NewA, 1, 35\\
Rangarajan F.V.N., Fabian A.C., Forman W.R., Jones C., 1995, MNRAS, 272, 665\\
Read A.M., 2003, MNRAS, 342, 715\\
Read A.M., Ponman T.J., 1998, MNRAS, 297, 143\\
Roberts T.P., Warwick R.S., 2000, MNRAS, 315, 98\\
Schweizer F., 1982, ApJ, 252, 455\\
Schweizer F., 1996, AJ, 111, 109\\
Schweizer F., Miller B.W., Whitmore B.C., Fall M.S.,1996, AJ 112, 1839\\
Sigurdsson S., Hernquist L., 1993, Nature, 364, 423\\
Stauffer J.R., 1982a, ApJ, 262, 66\\
Stauffer J.R., 1982b, ApJS, 50, 517\\
Stevens I.R., Hartwell J.M., 2003, MNRAS, 339, 280\\
Swartz D.A., Ghosh K.K., Tennant A.F., 2003, AAS, 202, 1109\\
Takizawa M., 1999, ApJ, 520, 514\\
Tozzi P. et al., 2001, ApJ, 562, 42\\
Webbink R.F., 1985, in Dynamics of Star Clusters, IAU Symposium 113, ed. J. Goodman \& P. Hut (Dordrecht: Reidel), 541\\
Whitmore B.C., Schweizer F., Leitherer C., Borne K., Robert C., 1993, AJ, 106, 1354\\
White R.E., Sarazin C.L., Kulkarni S.R., 2002, ApJ, 571L, 23 \\
Yun M.S., Hibbard J.E., 2001, ApJ, 550, 104\\
Zezas A., Fabbiano G., Rots A.H., Murray, S.S., 2002, ApJ, 142,239\\
Zwicky F., 1956, Ergebnisse der Exakten Naturwissenschaften, 29, 344\\

\end{document}